\documentclass[acmsmall]{acmart}
\usepackage{multirow}
\usepackage{longtable}
\AtBeginDocument{%
  }

\setcopyright{acmcopyright}
\copyrightyear{2023}
\acmYear{2023}
\acmDOI{XXXXXXX.XXXXXXX}

\acmJournal{CSUR}
\acmVolume{1}
\acmNumber{1}
\acmArticle{1}

\begin{document}

\title{A Survey on Off-chain Networks: Frameworks, Technologies, Solutions and Challenges}

\author{Xiaojie Wang} 
\email{wangxj@cqupt.edu.cn}

\author{Hanxue Li}
\email{S220132078@stu.cqupt.edu.cn}

\author{Ling Yi}
\email{yiling@cqupt.edu.cn}

\author{Zhaolong Ning}
\authornotemark[1]
\email{ningzl@cqupt.edu.cn}
\affiliation{%
  \institution{School of Communications and Information Engineering, Chongqing University of Posts and Telecommunications}
  \country{Chongqing, China}
}

\author{Song Guo}
\email{songguo@cse.ust.hk}
\affiliation{%
  \institution{Department of Computer Science and Engineering, The Hong Kong University of Science and Technology}
  \country{Hong Kong, China}
}

\author{Yan Zhang}
\email{yanzhang@ieee.org}

\affiliation{%
  \institution{Department of Informatics, University of Oslo}
  \country{Norway}
}

\renewcommand{\shortauthors}{Xiaojie Wang, et al.}

\begin{abstract}
 Blockchain has received increasing attention in academia and industry. However, the increasing transaction volumes and limited on-chain storage underscore scalability as a key challenge hindering the widespread adoption of blockchain.  Fortunately, off-chain networks that enable transactions outside the blockchain show promising potential to mitigate the scalability challenge. Off-chain solutions that address blockchain scalability hurdles, such as payment channel networks, facilitate secure and fast off-chain transactions, thus relieving the main chain's strain. In this article, we provide a comprehensive review of key technologies, solutions, and challenges of off-chain networks. First, we introduce the background of off-chain networks encompassing design motivation, framework, overview, and application scenarios. We then review the key issues and technologies associated with off-chain networks. Subsequently, we summarize the mainstream solutions for the corresponding key issues. Finally, we discuss some research challenges and open issues in this area.
\end{abstract}



\maketitle

\section{Introduction}
The development of new digital currencies is rapid, among which Bitcoin has gained significant influence and value~\cite{Lu2022CoinLayering}. At the same time, the arrival of Blockchain 3.0 allows assets to be tracked, controlled and transacted more securely on the blockchain, opening up a new world for the applications of blockchain technology in the financial field.

Despite the growing adoption of cryptocurrencies, they are poorly scalable. For example, Bitcoin records detail information about each transaction in the blockchain with the data size greater than 200 bytes~\cite{Kus2018ASurvey}, and the mining time for each block is controlled to be around 10 minutes. Although the maximum payment speed supported by the Bitcoin network is around 7 transactions per second~\cite{Xiao2020A}, and Ethereum processes 15 transactions per second~\cite{Zhang2021Boros}, it should be noted that the major challenge of scalability is the inefficiency of the underlying consensus protocol. Each transaction must undergo full consensus before it can be confirmed, the duration of which can last from a few minutes to several hours. As a result, blockchain is not suitable for latency-sensitive services.

Currently, there are significant shortcomings in the scalability of blockchain technology, mainly in the areas of block storage and transaction throughput. First, blockchain systems require storage spaces to increase with transaction amount and time, but on-chain storage spaces are limited and expensive. Furthermore, multiple transactions by blocks are accumulated and processed together, thus prolonging the transaction time. For large-scale transactions, slow on-chain computing can cause congestion, reducing transaction throughput. Therefore, existing blockchain systems can barely support a significant number of transactions. 

Off-chain networks that allow payment operations outside the blockchain are promising to solve scalability issues. Users can utilize payment channels to make multiple payments without having to submit each payment to the blockchain, and any party can close the payment channel by submitting the latest transaction information to the blockchain. The blockchain is only involved in creating and closing payment channels or in the event of disputes over the outcome. Off-chain networks are recognized as the key to off-chain payment, off-chain storage, and off-chain trusted computing integration, driving the development of off-chain transactions. 

Although blockchain has attracted widespread attention and facilitated people's life in areas, such as supply chain, entertainment, transportation, and information management~\cite{Xie2019A}, it still exhibits some shortcomings in terms of user privacy, storage costs, computing overhead, latency, and transaction speed. The emergence of off-chain networks is promising to relieve the above challenges. In recent years, off-chain networks are becoming a research hotspot, attracting a large number of researchers for in-depth exploration. Among them, \textbf{Payment Channel Networks (PCNs)} are widely used and are a proven and effective scaling solution. Currently, most researchers focus on the development of PCNs, but off-chain networks have not been thoroughly investigated.

\textbf{\textit{To the best of our knowledge, this is the first article dedicated to the study of off-chain networks.}} Therefore, we provide a concrete and comprehensive survey of off-chain networks. Specifically, the contributions of this article are as follows:

\begin{itemize}
	\item We first survey the background of off-chain networks, and then summarize the key issues in terms of off-chain data security, off-chain trust, transaction routing privacy, and off-chain transaction efficiency. After that, we investigate related technologies of off-chain networks.
	
    \item We summarize various solutions to solve key issues in off-chain networks, including privacy and security of data transmission, transaction throughput, data storage and sharing, and trusted off-chain data solutions. Subsequently, we provide the corresponding lessons learned.

	\item We discuss several open issues and future research directions for off-chain networks, including security and privacy protection of off-chain networks, transaction deadline of PCNs, lightweight off-chain storage, balance between transaction throughput and privacy risks.

\end{itemize}

The rest of this article is organized as follows. In Section \ref{two}, we introduce the background of off-chain networks. In Section \ref{three}, we discus key issues and promising technologies in off-chain networks. In Section~\ref{four}, we investigate relevant solutions in off-chain networks. Finally, we present challenges and future research directions in Section \ref{five}, and summarize this article in Section \ref{six}. 

\section{Background of Off-chain Networks}\label{two}

In this section, we first briefly analyze the motivation for off-chain network design. We then provide an overview of the off-chain network from its architecture and platform. Finally, we summarize potential application scenarios of off-chain networks.

\subsection{Motivation of Off-chain Network Design}

In this subsection, we present the motivation for off-chain network design in two aspects: from on-chain networks to off-chain networks and supported frameworks.

\subsubsection{From On-chain Networks to Off-chain Networks} 

Blockchain provides a decentralized, tamper-evident and verifiable ledger that can be used for recording and validating transactions. It has been widely applied in various fields, such as the \textbf{Internet of Vehicles (IoVs)}, healthcare, and financial services. However, blockchain suffers serious scalability issues, since the block size has capacity limitations and on-chain transactions may cause transaction congestion. Therefore, the throughput of transactions and the storage capacity of blockchain are obstacles to its widespread application. At the transaction level, the speed of on-chain transactions can vary depending on the number of transactions queued for processing, resulting in network congestion and expensive costs. In addition, in the case of complex smart contracts or other updated blockchain applications, transactions in on-chain networks may be more difficult to be managed than off-chain transactions.

Despite the widespread academic attention on blockchain applications, the high resource consumption of communication, computing, storage, and scalability limitations still require further resolution before applying it to real-world environments~\cite{Cao2023Blockchain}. In addition, blockchain relies on frequent communications among consensus nodes to reach consensus, bringing about high consumption of communication resources. In traditional blockchain networks, each consensus node needs to store a copy of the entire ledger~\cite{Arbabi2023A}, bringing huge storage pressure.

To address the scalability issues of on-chain networks, researchers have proposed various solutions. Among them, off-chain networks are considered as a promising approach. Specifically, off-chain networks allow transactions to be processed without being recorded on the main chain. In off-chain networks, transactions can be executed quickly without waiting for blockchain confirmations, thereby largely reducing latency. The nature of transactions in the off-chain network can alleviate the computing and storage burden, thereby enhancing transaction throughput and subsequently reducing transaction costs.

\subsubsection{Supported Frameworks} 

\indent The framework of off-chain networks is generally formed by PCNs and \textbf{Trusted Execution Environments (TEEs)}.

\indent \textbf{PCNs:} PCN is a blockchain-based system that enables high-efficient transactions between two parties without recording each transaction on the chain~\cite{Zhang2021RobustPay+}. Generally, some fees can be charged for forwarding transactions to encourage nodes to participate in forwarding. If two nodes do not have a direct connection, they can accomplish multi-hop payment relay through intermediate nodes. To ensure successful payment, the channel balance at each hop should be greater than the amount to be transacted.

\indent In short, PCNs allow participants to execute transactions in an off-chain manner by establishing payment channels~\cite{khalil2017revive}. Meanwhile, without frequent interaction with the slow blockchain, PCNs significantly reduce transaction latency and increase the blockchain throughput~\cite{Malavolt2017Concurrency}. PCN architecture includes multiple-signature addresses, fund locking, channel opening, transaction settlement, transaction circulation, and channel closure. First, users open a channel by depositing funds in on-chain. Once an off-chain payment channel is successfully established, users can make fast and low-cost transactions off-chain and only broadcast the final result to the blockchain. These processes significantly reduce transaction cost and time, while increasing transaction efficiency. Finally, either party can close the channel by submitting a transaction containing the latest balance message to the blockchain. After the transaction is completed, users can restart a new payment channel to make transactions. 

\indent \textbf{TEEs:} TEEs include secure areas of a device's hardware, software and memory for encryption and integrity protection. It aims to provide a secure environment for executing sensitive code and data, such as authentication and payment transactions, data encryption operations, and sensitive data processing. By offering a secure execution environment, TEE can effectively safeguard sensitive data from malicious attacks and tampering. Currently, TEE has been widely applied to enhance security and privacy of user data~\cite{Sebastian2019DER,ghafoorian2018thorough,kim2018sgx}. 

\indent TEE has been also utilized to mitigate scalability and confidentiality barriers of smart contracts. Even if adversaries compromise the operating system, they cannot learn about payment processes executed within TEEs. Through hardware protection and secure isolation techniques, TEEs ensure security, reliability, and efficiency of smart contract execution, so that they can reduce the burden on the blockchain network and off-chain networks' scalability.

\subsection {Overview of Off-chain Networks}
 
\indent In this subsection, we present an overview of off-chain networks from two aspects: architectures and platforms.

\subsubsection{Architectures}

\begin{figure}
	\centering
	\includegraphics[scale = 0.4]{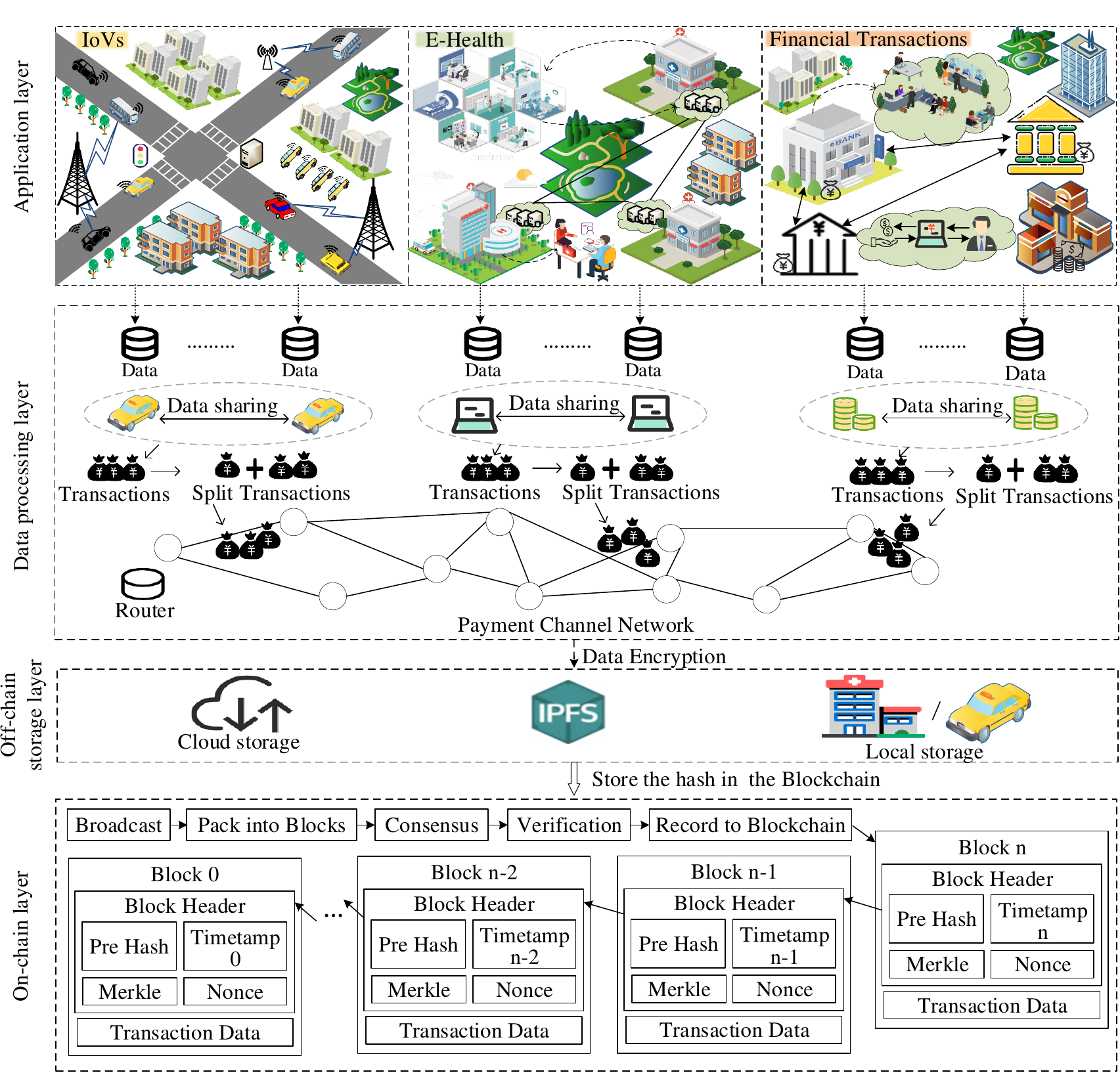}
	\caption{Architectures of off-chain blockchain.}
	\label{3:Architectures.}
	\Description{11}
	\vspace{-0.3cm}
\end{figure}

\indent The architecture of off-chain networks consists of the application layer, the data processing layer, and the off-chain storage layer, as shown in Figure~\ref{3:Architectures.}. Storing sensitive data directly on the blockchain can potentially lead to privacy leakage. Therefore, some researchers consider adopting off-chain storage methods (such as local storage, cloud storage, and \textbf{InterPlanetary File System (IPFS)} storage), and storing only the hash value or data identifier on the blockchain to ensure privacy and security. First, the application layer transmits the collected data to the data processing layer for further processing. Then, the data in the processing layer is encrypted and stored in the off-chain storage layer. Finally, the hash of the encrypted data is uploaded to the main chain.

\indent The data processing layer of off-chain networks includes data processing, transaction division and PCNs. The data sharing among the parties, including vehicles, hospitals, and banks, occurs outside the main blockchain without the need for global consensus. Large-scale off-chain transactions are always preferred to be divided into several smaller transactions and then forwarded through different paths. However, during this process, users need to pre-deposit enough funds to prompt successful transactions. Furthermore, processed data is stored off-chain rather than being directly stored on the blockchain network, thus alleviating the burden on the blockchain and contributing to the maintenance of transaction privacy. 

\subsubsection{Platforms}

\indent Typical off-chain network platforms include Bitcoin Lightning Network~\cite{Kurt2023A}, Ethereum Raiden Network~\cite{3453161}, and Ripple Coin~\cite{Zhang2021RobustPay+}. They are reckoned to have the advantage of improving off-chain transaction speed and throughput while reducing transaction latency and cost. The Lightning Network for Bitcoin and the Raiden Network for Ethereum are the most widely deployed off-chain PCNs in practice~\cite{Du2022Anti-Collusion,Egger2019Atomic}. The Ripple Network is currently the only off-chain network with publicly available transaction data. 

\indent The Bitcoin Lightning Network, which enables off-chain payments among Bitcoin participants, relies on penalties rather than time locks to encourage honest behaviour. The network enables fast, low-cost, and highly reliable peer-to-peer payments by establishing bidirectional payment channels. It is a well-known PCN project for Bitcoin using \textbf{Hashed TimeLock Contract (HTLC)} to complete off-chain payments while ensuring security~\cite{Chen2022Multi}. 

\indent The Raiden Network is similar to the Lightning Network, and achieves efficient peer-to-peer payments by establishing bidirectional payment channels. Unlike the Lightning Network, anyone can create and manage the Raiden network’s unlimited payment channels, dramatically increasing the network’s scalability. Furthermore, Ripple Network achieves payments by establishing direct connections and trust relationships. Simultaneously, it offers fast transactions, low fees, high scalability, and strong security as well as reliability.

\subsection{Application Scenarios}

\indent Off-chain networks can meet data storage and processing demands in different application scenarios such as E-health, IoVs, and financial transaction applications. Simultaneously, PCNs, as one kind of the most promising off-chain transaction networks, also provide off-chain transaction channels for \textbf{Electronic Medical Records (EMRs)}, IoVs and financial transactions, enabling efficient and secure data transmission among devices. Based on that, we describe the corresponding application scenarios from three aspects: E-health, IoV, and financial transactions.

\subsubsection{E-health}

\indent With the rapid development of information technology, medical services are gradually moving towards digitalization and informatization. Patient's health conditions, medical treatments, and other relevant information are digitally recorded by EMRs. It has shown significant benefits in reducing healthcare costs, improving clinical quality and enhancing disease surveillance. However, most hospitals currently use centralised storage models for sensitive information, which comes with many disadvantages, such as a single point of failure, data leakage, and privacy invasion.

\indent The off-chain networks ensure privacy and security of e-health through distributed storage, encryption algorithms, smart contracts and other technologies. Since the blockchain is difficult to store a large number of medical records, a distributed off-chain storage system is required. Authors in~\cite{Azbeg2022Access,Liang2022PDPChain} adopt IPFS for off-chain storage of medical records, satisfying the storage requirements of blockchain and leveraging encryption technology to achieve patient privacy and security. Authors in~\cite{Wu2022Blockchain} use IPFS to store published EMR data, and design a local differential privacy-based policy and a role-based access control approach to achieve privacy protection for data sharing.

\subsubsection{IoVs}

\indent IoV has become a vision for intelligent transportation systems with great potential~\cite{Wang2023Blockchain,Zhang2016Social}, since traditional traffic management schemes have the disadvantage of low efficiency and result in an inconvenient travel experience for users. Data sharing is a key feature of IoVs. Vehicles in the network share and access data with other participants to provide services, such as traffic management, accident avoidance and improved driving experience~\cite{Alladi2022A}. 

\indent In recent years, a large number of off-chain based IoV services have been provided. For the application of off-chain networks in IoV, authors in~\cite{Yuan2023TRUCON} overcome the problem of channel congestion caused by extensive data sharing of vehicles, and the off-chain proportion effectively controls the congestion state. Different from~\cite{Yuan2023TRUCON} without considering privacy issues, electric vehicle generates pseudonymous addresses off-chain and uses them to interact with contracts for authentication in~\cite{Gabay2020Privacy}. The authors combine zero-knowledge proofs and smart contract techniques to protect the privacy of electric vehicles during charging. 

\indent Additionally, crowdsensing systems are promising to access sensory data from ubiquitous mobile devices, providing significant value for applications, such as intelligent transportation~\cite{Zheng2017Privacy}. In the field of IoV, vehicle data contain a lot of user information, such as trajectories, traffic information and private data. The leakage of sensitive information has a significant negative impact on users. 

\subsubsection{Financial Transaction}

\indent Off-chain networks can be used for identity verification in financial transactions, ensuring trustworthiness and security of transactions. They can offer efficient data processing and storage capabilities to support large-scale financial transactions.

\indent The blockchain ledger is open and transparent, and attackers can analyze the ledger data to obtain sensitive information, posing a threat to users' transaction privacy. For example, in Internet shopping chains, attackers can infer users' purchasing power, income levels, consumption habits, and deduce their real identities by analyzing their transaction behaviors and transaction amounts. Authors in~\cite{Fan2020Hybrid} apply payment channels to resource transactions and propose a blockchain-based hybrid resource system. They utilize payment channel technology to achieve trustworthy, fast, and low-cost payment transactions, but fails to meet the security requirements.

\indent It is known that financial transactions involve large amounts of sensitive information, and issues such as the difficulty of data sharing and potential data security risks are widespread. To address these concerns, privacy protection technologies and off-chain networks are combined to ensure that transaction information is only accessible to authorized users. Meanwhile, off-chain networks store transaction data from different financial institutions in the off-chain storage (i.e., IPFS, cloud servers, and local servers), alleviating the pressure on on-chain storage and enhancing blockchain performance.

\section{Issues and Technologies of Off-chain Networks}\label{three}

\indent Although off-chain transactions can improve scalability and transaction speeds of blockchain networks, many issues remain to be solved. This section first analyzes the key issues, and then introduces the related technologies of off-chain networks in Table~\ref{table1}.

\subsection{Issues of Off-chain Networks}\label{Issues}

\indent As yet, off-chain networks still struggle with off-chain data security, off-chain trust, transaction routing privacy, and off-chain transaction efficiency. In the following, we discuss them in detail.

\subsubsection{Off-chain Data Security}

\indent Security is the major challenge in off-chain networks. Specifically, exposing users' privacy and other sensitive information can cause a series of security risks. Although distributed ledger systems provide high security for on-chain data~\cite{Ning2022SmartGrids}, privacy and security of off-chain data are not guaranteed. 

\indent When off-chain transaction is processed, the instantaneous channel is not publicly announced in PCNs to maintain data security and privacy. Therefore, the transaction sender needs to iteratively probe the candidate path for available balances before sending the transaction, but frequent channel probes may violate privacy~\cite{Qiu2022A}. Additionally, during the transaction process, it is important to encourage participants to engage in data sharing. Dishonest participants providing malicious data can increase the risks of data leakage and tampering. Therefore, ensuring data security of off-chain networks remains a crucial concern. 

\subsubsection{Off-chain Trust}

\indent It refers to trust established through mechanisms and channels outside of blockchain. In contrast, trust in on-chain transactions relies on cryptographic protocols and consensus mechanisms, thereby ensuring transaction validity and security. However, some studies have shown that trust in blockchain is limited to the on-chain and does not extend to the off-chain environment, thus creating a ``trust gap'' with the physical off-chain environment~\cite{Liu2022Extending}. 

\indent  In public systems, shared data may become public, leading to a lack of trustworthiness. In PCNs, some untrusted users participating in transactions may refuse to cooperate and tamper with transactions, resulting in transaction failures or  compromised transaction outcomes. Completing off-chain transactions requires submitting the transaction results for confirmation on the blockchain. However, during this process, there is a risk of result tampering or other issues that can impact the effectiveness of transactions.

\subsubsection{Transaction Routing Privacy}

\indent Since different routes charge different fees, the channels on the path need to have sufficient available funds. Therefore, the available balance of each channel should be greater than the amount being sent. Moreover, when data transactions are performed off-chain, routing in PCNs is highly sensitive to the dynamics of channel balancing due to the consuming nature of transaction forwarding. Some researchers have made progress on routing schemes for PCNs, but their privacy is still not well protected~\cite{Roos2018Settling,prihodko2016flare,PengW2019Flash}.

\indent PCNs serve as a form of off-chain transaction networks and allow two nodes to complete multiple payments without resorting to on-chain operations. When processing transactions off-chain, two nodes can open a channel by depositing an initial amount, and later execute off-chain transactions by updating the available balance in the channel. Since each transaction involves multiple participants, it may expose information about the transaction routing, thereby compromising information privacy. To address the above issues, authors in~\cite{pietrzak2021lightpir} propose a privacy-preserving route discovery mechanism, which solves the privacy leakage problem in PCNs. However, their implementations are based on the assumption of static networks, without considering dynamic routing. Most studies only considered privacy and static routing separately, since balancing privacy issues and dynamic routing encountered in off-chain transactions is very challenging.

\subsubsection{Off-chain Transaction Efficiency}

\indent Payment networks are composed of multiple payment channels, supporting off-chain transactions with multiple hops. Simultaneously, a collection of bidirectional payment channels forms a PCN. 

\indent The transaction efficiency in PCNs is influenced by factors such as transaction rates, deposited funds, channel congestion, and transaction deadline. If the transaction rate in one direction of a payment channel is higher than that in the other direction, the channel may become unbalanced. This can result in the depletion of funds in many channels, thus affecting the transaction efficiency in PCNs. Existing routing protocols attempt to atomically and instantaneously route each incoming transaction. However, for a large transaction, it may fail if a path with sufficient funding to the destination cannot be found. In PCNs, the increasing number of transactions and the sharing of payment channels for concurrent transactions may cause channel congestion. When payment channels are congested, transactions with longer deadlines may be processed earlier than those with shorter deadlines, resulting in the failure of transactions close to the deadline.

\indent  Authors in~\cite{yu2018coinexpress} propose a fast routing mechanism to solve the issue of partial payment channel congestion. However, the mechanism incurs high overhead when the entire network is congested. Therefore, the issue of improving the transaction efficiency in PCNs remains worthy of further research.

\subsection{Technologies for Off-chain Networks}

\indent This subsection details security and privacy based technologies, intelligent off-chain networking and routing technologies, off-chain edge computing technologies, and off-chain transaction scheduling technologies in the latest research.

\begin{table*}[htbp]
	\centering
	\caption{\centering Technologies for Off-chain Networks}
	\label{table1}
	\centering
	\linespread{1.1}\selectfont
	\begin{tabular}{|c|c|c|}
		\hline
		
		\multirow{2}*{\parbox[c]{17mm}{\centering\textbf{Focus}}} & 
		\multirow{2}*{\parbox[c]{6mm}{\centering\textbf{Ref.}}} & 
		\multirow{2}*{\parbox[c]{117mm}{\centering\textbf{Description}}} \\
		
		&& \\
		
		\hline
		\multirow{12}*{\parbox[t]{17mm}{\centering Security and privacy}} &  \multirow{2}*{\centering ~\cite{li2022efficient}}  & {\parbox[t][8mm]{117mm}{ Design on-chain ledger and off-chain storage management model, thereby saving on-chain space and preserving data security}}\\
		
		\cline{2-3}
		& \multirow{2}*{\centering ~\cite{lin2021model}}  & {\parbox[t][8mm]{117mm}{ Combine on-chain and off-chain collaboration mechanisms with reputation assessment models to alleviate on-chain storage pressure and enhance data security }}\\
		
		\cline{2-3}
		& \multirow{2}*{\centering ~\cite{Ma2021Research}}  & {\parbox[t][8mm]{117mm}{ Design a security mechanism that combines on- and off-chain data transmission and encryption algorithms to achieve collaborative on-chain and off-chain protection}}\\
		
		\cline{2-3}
		& \multirow{2}*{\centering ~\cite{Cai2019Towards}}  & {\parbox[t][8mm]{117mm}{ Combine lightweight encryption and off-chain private data validity checking to protect data privacy }}\\
		
		\cline{2-3}
		& \multirow{2}*{\centering ~\cite{liu2021blockchain}}  & {\parbox[t][8mm]{117mm}{ Combine encryption technologies with collaborative on-chain and off-chain computing to enhance data security}}\\

		\cline{2-3}
		& \multirow{2}*{\centering ~\cite{Xue2022Blockchain}}  & {\parbox[t][8mm]{117mm}{ Design a key update mechanism by using identity-based hierarchical encryption to improve data transmission security }}\\
		\hline

		\multirow{11}*{\parbox[t]{17mm}{\centering Intelligent off-chain networking and routing}} &\multirow{2}*{\centering ~\cite{Qiu2022A}}  & {\parbox[t][8mm]{117mm}{ Propose a privacy-aware high-throughput routing algorithm based on DRL that considers both privacy preservation and long-term throughput }}\\

		\cline{2-3}
		& \multirow{2}*{\centering ~\cite{He2022Bift}}  & {\parbox[t][8mm]{117mm}{ Combine FL, off-chain IPFS and blockchain to support on-chain and off-chain data sharing and protect user privacy }}\\
		
		\cline{2-3}
		& \multirow{2}*{\centering ~\cite{Wang2023AI}}  & {\parbox[t][8mm]{117mm}{ Combine FL, blockchain, and IPFS to ensure trustworthy training of anomaly detection models and alleviate on-chain storage pressure }}\\
		
		\cline{2-3}
		& \multirow{1}*{\centering ~\cite{Ye2022FLasaS}}  & {\parbox[t][4mm]{117mm}{ Combine FL, off-chain storage and trusted mechanisms to achieve privacy protection }}\\
		
		\cline{2-3}
		& \multirow{2}*{\centering ~\cite{Li2022Compact}}  & {\parbox[t][8mm]{117mm}{ Propose a compact DRL algorithm to learn the joint dynamic and lightweight routing policy for maximizing long-term transaction efficiency }}\\

		\cline{2-3}
		& \multirow{2}*{\centering ~\cite{Ning2022Blockchain}}  & {\parbox[t][8mm]{117mm}{ Use DRL algorithms to achieve the trade-off between blockchain latency and security by legally selecting transactions from transaction pools }}\\
		\hline
		
		\multirow{8}*{\parbox[t]{17mm}{\centering Off-chain edge computing}} & \multirow{2}*{\centering ~\cite{wang2021lifesaving}}  & {\parbox[t][8mm]{117mm}{ Design an optimal computing offloading mechanism, which offloads storage and computing tasks from on-chain to off-chain vehicles to alleviate on-chain pressure}}\\

		\cline{2-3}
		& \multirow{2}*{\centering ~\cite{Wang2022Mean}}  & {\parbox[t][8mm]{117mm}{ Design a learning-based off-chain computing offloading algorithm to maximize the long-term utility of miners }}\\

		\cline{2-3}
		& \multirow{2}*{\centering ~\cite{Kang2019Blockchain}}  & {\parbox[t][8mm]{117mm}{ Combine blockchain, vehicular edge computing and networks to achieve secure data storage and sharing }}\\

		\cline{2-3}
		& \multirow{2}*{\centering ~\cite{zhang2022blockchain}}  & {\parbox[t][8mm]{117mm}{ Utilize PCN-based blockchain to enable data sharing among IoT and introduce transaction splitting scheme to improve the transaction success rate }}\\
		\hline
		
		\multirow{11}*{\parbox[t]{17mm}{\centering  Off-chain transaction scheduling}}    & \multirow{2}*{\centering ~\cite{yu2018coinexpress}}  & {\parbox[t][8mm]{117mm}{ Propose a novel distributed dynamic routing mechanism to find "express lanes" for cryptocurrency-based digital payments in PCNs}}\\
		
		\cline{2-3}
		& \multirow{1}*{\centering ~\cite{Zhang2021Robustpay}}  & {\parbox[t][4mm]{117mm}{ Propose a distributed robust payment routing protocol to resist transaction failures}}\\

		\cline{2-3}
		& \multirow{2}*{\centering ~\cite{Luo2022Learning}}  & {\parbox[t][8mm]{117mm}{ Propose priority-aware PCNs to achieve efficient transaction scheduling and select forwarding cost as a priority identifier}} \\
		
		\cline{2-3}
		& \multirow{2}*{\centering ~\cite{Papadis2022Single}}  & {\parbox[t][8mm]{117mm}{ Design single-hop transaction scheduling by considering transaction deadlines to maximize throughput}}\\
		
		\cline{2-3}	   
		& \multirow{2}*{\centering ~\cite{Wang2022DPCN}}  & {\parbox[t][8mm]{117mm}{ Propose a deadline-aware PCN framework that considers transaction deadlines to improve the transaction success ratio}}\\
		
		\cline{2-3}
		& \multirow{2}*{\centering ~\cite{sivaraman2020high}}  & {\parbox[t][8mm]{117mm}{ Propose multi-path transmission protocols and congestion control algorithms to promote channel balancing and improve throughput in PCNs}}\\
		
		\hline

	\end{tabular}
\end{table*}


\subsubsection{Security and Privacy based Technologies}

\indent Off-chain networks typically use on-chain certificate and off-chain storage, data collaboration technologies for on-chain and off-chain, and encryption technologies to ensure data privacy and security.

\textbf{On-chain Certificate and Off-chain Storage:} With the increasing storage and recording of data information in electronic form, the issue of data privacy and security has become increasingly prominent due to the lack of trust among departments and the risks of data leakage. As a result, many data owners are unwilling to share their data. Additionally, directly putting data on the blockchain can result in high storage costs, excessive network consumption, and inadequate privacy protection. Some researchers adopt off-chain storage methods (i.e., local storage, cloud storage, and IPFS storage), and store only the hash value or the data identifier on the blockchain to ensure privacy and security~\cite{10.1145/3530813}.

\indent The on-chain certification, off-chain storage and transmission methods can also solve the above problems. For example, only requests and responses for shared data are recorded on the blockchain, while the actual data sharing is accomplished through off-chain transmission. It protects data privacy and reduces the burden on the blockchain, thereby enhancing system throughput. Authors in~\cite{wang2021lifesaving} proposes a vehicular fog computing mechanism based on off-chain storage. It utilizes ground vehicles as mobile fog nodes, and offloads extensive computation tasks of UAVs to off-chain ground vehicles for storage and processing. Merely depositing data pointers to the on-chain can alleviate its heavy burden. Authors in~\cite{li2022efficient} propose an on-chain ledger and off-chain storage model to manage electronic cases in healthcare delivery systems, uploading the corresponding indexes to the on-chain. However, the actual EMR data is stored on off-chain servers, thus saving on-chain space and protecting data security. 

\indent \textbf{Data Collaboration Technologies for On-chain and Off-chain:} The size of blocks is fixed, and storage space becomes limited with the increasing number of transactions. Therefore, the data collaboration technology that combines the on-chain and the off-chain can achieve efficient data processing and storage. It integrates searchable attribute encryption technology to build a secure and controllable data collaboration architecture, which ensures data privacy and security in the off-chain. The retrieval and analysis of on-chain data is realized by means of block indexing. 

\indent The study in~\cite{lin2021model} constructs a secure and efficient model training mechanism based on on-chain and off-chain collaboration. It stores model parameter off-chain, reducing storage pressure on edge nodes. Additionally, the original data does not have to be uploaded to centralized servers, which helps protect user privacy and decrease network overhead. Authors in~\cite{Ma2021Research} propose a power data collection scheme that divides data transmission into on-chain and off-chain processes. On-chain and off-chain collaborative protection mechanisms are developed based on encryption algorithms, aiming to protect off-chain data. Authors in~\cite{Cai2019Towards} design on-chain and off-chain co-computing in secure processing to reduce on-chain overhead. They use lightweight cryptography for data encryption and propose an off-chain validity checking strategy for private data.

\indent \textbf{Encryption Technologies for Off-chain Networks:} They refer to encrypting and protecting data in off-chain environments, aiming to ensure data confidentiality, integrity and authentication. In off-chain networks, common encryption techniques for privacy protection include symmetric and asymmetric encryptions. Among them, symmetric encryption uses the same key to encrypt and decrypt data, while asymmetric encryption uses both public and private keys.

\indent Encryption technology can prevent the leakage of private data in the off-chain environment, since attackers cannot obtain the plaintext data even if the ciphertext is compromised. The study in~\cite{liu2021blockchain} proposes an attribute encryption based anti-forgery sharing model. Different from the traditional attribute encryption algorithm, they introduce random parameters as identity keys and encryption key pairs to encrypt private keys. This improvement enables keys to undergo off-chain computation across multiple nodes and achieves on-chain storage. To efficiently update the keys of multiple users in case of key exposure, authors in~\cite{Xue2022Blockchain} use digital keywords and time validity to encrypt different types of data. They construct an efficient key update mechanism by identity-based hierarchical encryption, where data owners need to update their keys periodically to improve the security of data transmission.

\subsubsection{Intelligent Off-chain Networking and Routing Technologies} Computational tasks can be offloaded from the blockchain to off-chain networks, with the purpose of improving blockchain performance and throughput. Intelligent off-chain networks can provide great flexibility in protecting data privacy. \textbf{Federated Learning (FL)} and \textbf{Deep Reinforcement Learning (DRL)} based technologies are illustrated to support intelligent off-chain networks, enabling efficient data processing and routing.

\textbf{FL enabled Off-chain Networking:} FL is a distributed \textbf{Machine Learning (ML)} technique that exhibits strong performance in mitigating user privacy concerns~\cite{9565851,9918062,9460016}. It enables model training without the need for data sharing~\cite{Wang2019Adaptive}. Nonetheless, despite the manifold advantages of FL, its decentralized nature also introduces certain security vulnerabilities. By storing and sharing models on the blockchain, the decentralized security issues faced by FL can be effectively addressed. Furthermore, in off-chain environments, FL involves conducting model training on local devices and storing model parameters in IPFS. This approach effectively eliminates delays caused by blockchain transaction confirmations and ensures data privacy. Currently, many research efforts are devoted to the study of FL to facilitate the development of off-chain networks. 

\indent FL is adopted in intelligent healthcare to establish collaborative healthcare environments among hospitals. In this way, patients do not need to share data among healthcare organizations, thus expediting diagnosis and treatment, and reducing the risk of user privacy leakage~\cite{9415623}. Authors in~\cite{9928220} propose a platform-free proof of FL scheme, which involves uploading the trained model data to an off-chain IPFS platform for storage, thereby alleviating the storage pressure on the blockchain. Authors in~\cite{He2022Bift} combine FL, blockchain, and IPFS to preserve the privacy of connected and autonomous vehicles. They employ IPFS for data storage to support efficient exchange of on-chain and off-chain data. The study uses distributed IPFS off-chain storage to provide an efficient and stable data-sharing system and simultaneously reduces the load on the blockchain. The authors in~\cite{Wang2023AI} combine blockchain and FL to ensure trustworthy training of trajectory anomaly detection models. It aims to realize global model fusion through on-chain and off-chain data exchange. The authors store the model parameters of the training process to IPFS servers to reserve blockchain storage resources. The authors in~\cite{Ye2022FLasaS} use distributed ledger technology to support FL as a service. It aims to achieve trusted on-chain and off-chain data management by using trusted mechanisms, off-chain storage of raw data, and on-chain storage of hashes.

\begin{figure}
	\centering
	\includegraphics[width=0.7\textwidth]{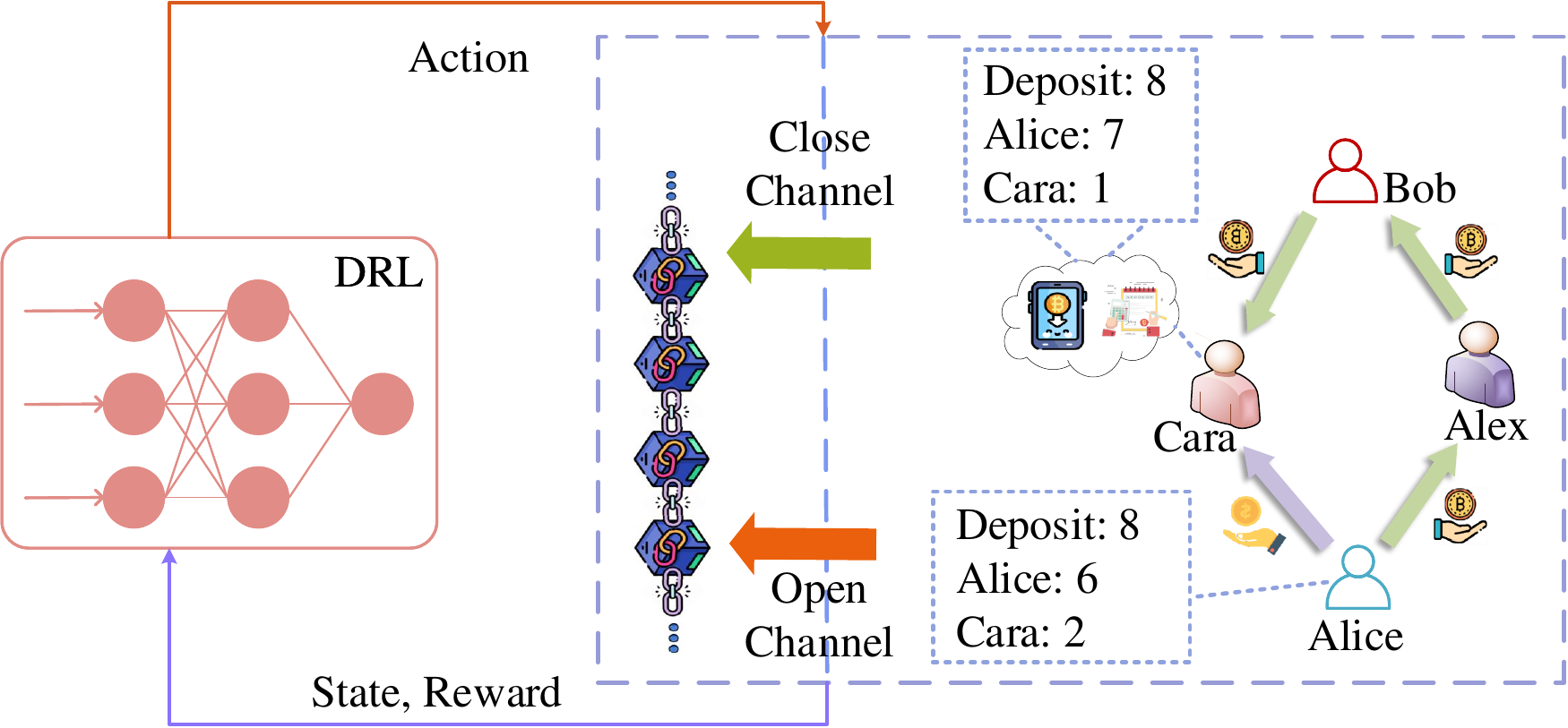}
	\caption{DRL model for off-chain routing.}
	\label{4:DRL.}
	\Description{11}
	\vspace{-0.5cm}
\end{figure}

\indent \textbf{DRL enabled Off-chain Routing:} DRL algorithms can learn the optimal routing paths based on factors, such as network congestion, latency, and reliability. As shown in Figure \ref{4:DRL.}, in the context of off-chain networks, DRL can train agents through interactions with the environment and learn optimal off-chain transaction actions via reward or punishment feedback, thereby enabling successful off-chain transactions and adapting to changing environments.

\indent There are two protocols for network routing in off-chain networks, i.e., static routing and dynamic routing. Authors in~\cite{Roos2018Settling} consider embedding-based routing, which assigns coordinates to nodes and finds ways to reduce the average path length. Authors in~\cite{malavolta2016silentwhispers} focus on landmark-centered routing, which requires all paths to pass through the landmark. However, they both belong to static routing~\cite{Roos2018Settling,malavolta2016silentwhispers}. In off-chain networks, dynamic routing focuses on dynamic channel balancing. Authors in~\cite{sivaraman2020high} apply dynamic routing in off-chain networks, achieving better performance than static routing protocols. Meanwhile, the authors utilize routing solutions based on packet transactions and multi-path transport protocols to achieve high throughput routing. 

\indent ML has efficient inference and decision-making capabilities in analyzing various data types collected by edge devices~\cite{chen2021distributed}. In PCNs, DRL can enable intelligent data routing in large-scale networks. The authors in~\cite{Li2022Compact} employ DRL algorithm to learn path selection policies, aiming to address dynamic off-chain routing issues in PCNs. The authors in~\cite{Qiu2022A} design a privacy-aware routing algorithm based on DRL that uses instantaneous throughput as a reward, and focuses on maximizing cumulative rewards to achieve long-term high throughput. Authors in~\cite{Ning2022Blockchain} construct a blockchain-enabled crowdsensing framework that collectively considers user data, utility, and system latency. They utilize vehicle crowdsensing techniques for off-chain ground vehicle data collection, and employ the DRL algorithm to legitimately select active miners and transactions from \textbf{Road Side Units (RSUs)}, to achieve an effective trade-off between latency and data security.

\subsubsection{Off-chain Edge Computing Technologies}

\begin{figure}
	\centering
	\includegraphics[width=\textwidth]{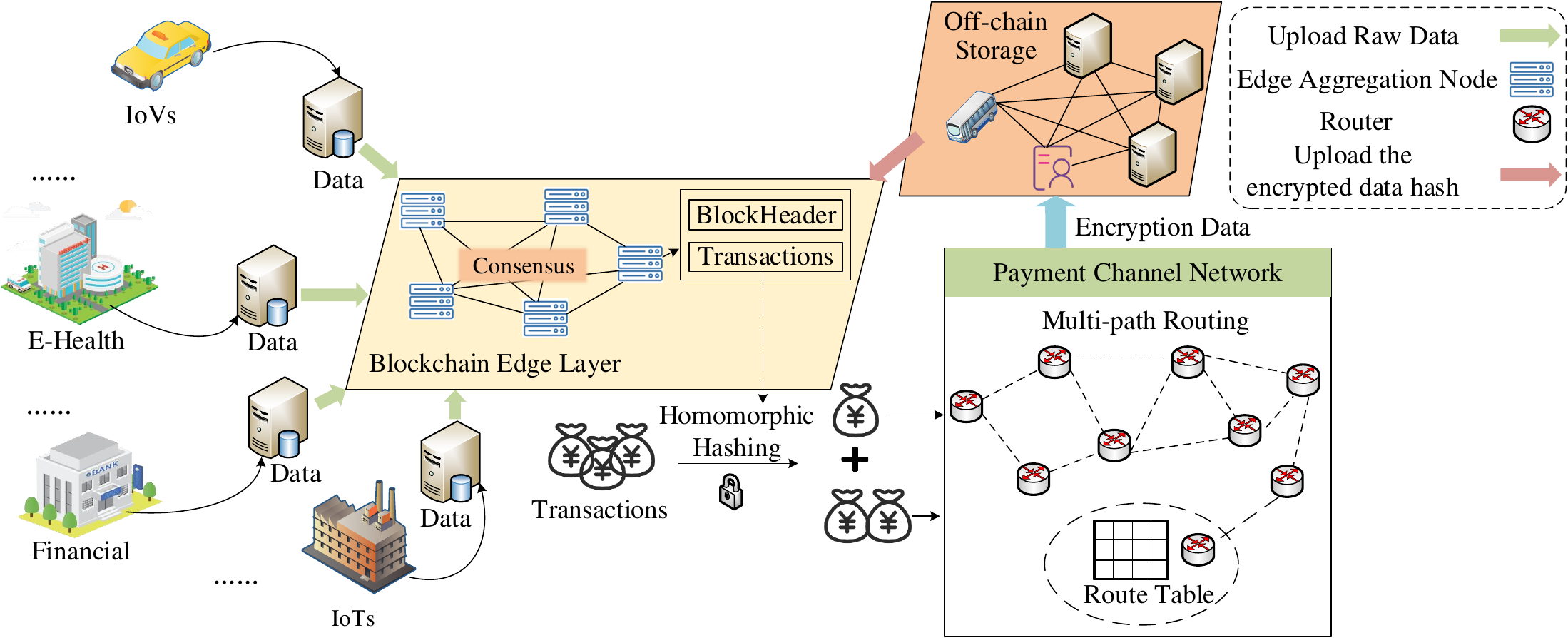}
	\caption{Off-chain edge computing.}
	\label{5:EC.}
	\Description{11}
		\vspace{-0.5cm}
\end{figure}

\indent In blockchain networks, computation offloading allows terminal devices to offload decision-making and data-processing tasks to off-chain edge nodes. This helps alleviate the computational burden on terminal devices and supports complex computational tasks of the blockchain. Additionally, off-chain data sharing among participants improves data availability. As shown in Figure \ref{5:EC.}, in scenarios like the \textbf{Internet of Things (IoT)}, E-Health, and financial institutions, raw data is shared and stored off-chain, then uploaded to the blockchain edge for processing. Transactions at edge consensus nodes require users to deposit funds to open transaction channels. Subsequently, transactions are divided into small units and forwarded through multi-hop routing. Finally, data is encrypted and stored off-chain to expand the storage capacity of blockchain.

\textbf{Computation Offloading from On-chain to Off-chain:} Computation offloading allows computational tasks to be offloaded from local devices to servers in off-chain networks, to reduce device energy consumption, decrease data transmission latency, and enhance network efficiency.

\indent Complex computational tasks can be offloaded from the blockchain network to off-chain networks for processing~\cite{asheralieva2019learning}. Subsequently, results are submitted to the on-chain for verification and validation. This approach reduces the burden of on-chain computation and improves the overall network efficiency. Authors in~\cite{wang2021lifesaving} utilize the idle computing and storage resources of off-chain ground vehicles to design an \textbf{Reinforcement Learning (RL)}-based optimal computation offloading mechanism. This mechanism offloads computing and storage tasks of on-chain UAVs to ground vehicles. Off-chain storage and computing at the network edge compensate for storage and computing resources on the blockchain. 

\indent Currently, most researchers only consider edge servers managed by one service provider to support the mining process. Instead, authors in~\cite{Wang2022Mean} investigate scenarios regarding multiple service providers. To address this problem, they first build a Markov game by considering the miner's task completion latency and block size, and subsequently design a learning-based off-chain computation offloading algorithm to achieve the goal of maximizing the long-term utility of miners.

\indent \textbf{Off-chain Data Sharing:} In recent years, secure and efficient data sharing services are provided across various fields. By using mobile edge computing techniques, authors in~\cite{Kang2019Blockchain} propose a secure and trustworthy data-sharing mechanism in the IoV, where mobile edge nodes run the hybrid proof of work and proof-of-storage consensus protocol for maintenance. The study in~\cite{Chai2021Hierarchical} presents a decentralized model training scheme in edge computing-enabled IoVs in urban areas for collaborative environment sensing, data processing, and knowledge sharing among vehicles. 

\indent However, these solutions are not applicable to disaster scenarios. Fortunately, UAVs can provide reliable emergency networks for disaster areas when ground communication infrastructure fails. However, there are potential security threats to UAVs during data transmission. A secure and efficient data sharing scheme for UAV-aided disaster relief networks is designed in~\cite{wang2021lifesaving} to ensure secure data sharing. The reluctance to share data among nodes cannot meet the high-frequency data sharing needs among IoT devices. Authors in~\cite{zhang2022blockchain} propose an IoT data sharing method by using the PCN-extended blockchain to support high-frequency data interchange. The authors develop a transaction splitting scheme based on homomorphic hashing, which solves the problem of low transaction success rates caused by deposit restrictions in PCNs. Additionally, they propose a multi-path routing scheme based on multi-point relays, which generates multiple routing paths to improve transaction success rates.

\subsubsection{Off-chain Transaction Scheduling Technologies}

\indent In PCNs, off-chain transactions are efficiently processed through specific scheduling strategies. Participants complete off-chain transactions through payment channels, avoiding transaction confirmation time and fees on the blockchain. Furthermore, off-chain transaction scheduling technologies play a key role in PCNs. Through appropriate scheduling and routing, it can enhance the efficiency and performance of PCNs. 

\indent \textbf{Priority Scheduling:} Off-chain transactions can reduce the load on the blockchain, since users conducting transactions off-chain do not need to submit each transaction to the blockchain. Transactions are scheduled based on different priorities, and each hop along the transaction path requires the payment of corresponding forwarding fees. In PCNs, transaction priority scheduling often considers factors such as fees, transaction sizes, and transaction latency.

\indent Compared to traditional transactions, the transaction priority scheduling based on off-chain networks can shorten the transaction time, reduce channel congestion, and decrease the total transaction forwarding costs. Existing research focuses on the routing issues in PCNs~\cite{Roos2018Settling,prihodko2016flare,malavolta2016silentwhispers,PengW2019Flash}. However, the requirement for transaction atomicity comes at the cost of delay, which affects the user experience of delay-sensitive applications. Authors in~\cite{Papadis2022Single} investigate single-hop transaction scheduling issues by considering the transaction deadline. The study in~\cite{Wang2022DPCN} designs a deadline-aware PCN framework through the collaboration of dynamic transaction segmentation based on delay-sensitive transaction scheduling and congestion avoidance algorithms. When conducting multi-hop routing for transactions, the sender needs to pay fees to the intermediate nodes for transaction forwarding~\cite{Roos2018Settling}. The priority payment channel is the one with the least channel cost calculated by the transaction amount. In addition, to achieve efficient transaction scheduling, authors in~\cite{Luo2022Learning} propose a priority-aware PCN framework. They choose the forwarding cost as the priority identifier and aim to achieve high throughput by adjusting fees paid to each hop.

\indent \textbf{Congestion Control:} PCNs have great potential to improve blockchain throughput, but the growing number of transactions and the sharing of payment channels by concurrent transactions may cause channel congestion~\cite{Luo2022Learning}. When dealing with mass transactions, single-path routing algorithms may fail due to the lack of high capacity. Such algorithms can also lead to routing centrality and channel congestion resulting from the biased selection of high-capacity channels~\cite{prihodko2016flare}. Multi-path routing algorithms can effectively reduce channel congestion by splitting transactions.

\indent Many blockchain providers use PCNs to increase their transaction throughput. However, concurrent PCNs suffer from serious congestion problems. To address this issue, researchers have made efforts in various aspects. For example, authors in~\cite{yu2018coinexpress} propose a fast routing mechanism for payment channels to solve congestion issues. This work imposes a large overhead when the entire network is mostly congested. Similar to~\cite{yu2018coinexpress} which can generate additional transaction load, authors in~\cite{Zhang2021Robustpay} propose a multi-path scheduling method that sends transactions to different routes to counter transaction failure. The study in~\cite{sivaraman2020high} introduces a multi-path transaction flow scheduling method that dynamically describes channel congestion levels. It divides transactions into several units sent at different rates through different paths. By controlling the number of transaction units allocated to each path, the utilization rate of payment channel funds is improved. Since transaction delays are high in congested networks, authors in~\cite{Wang2022DPCN} study congestion avoidance algorithms, and congestion marking is performed for transactions with long queues.

\section{Solutions of Off-chain Networks}\label{four}

\indent The large storage capacity and flexibility of off-chain networks make them suitable for large-scale data storage and processing, while their high real-time, low latency, and suitability for fast transactions also attract attention. However, off-chain networks still face issues of data security, trust, transaction privacy, and efficiency in different scenarios. In the face of various problems mentioned in subsection~\ref{Issues}, this section illustrates several solutions to address these issues. In the following, we summarize the relevant research on these solutions in Tables~\ref{table3} and~\ref{table4}  and give lessons learned.  

\subsection{Privacy and Security of Data Transmission}

\begin{figure}
	\centering
	\includegraphics[scale = 0.55]{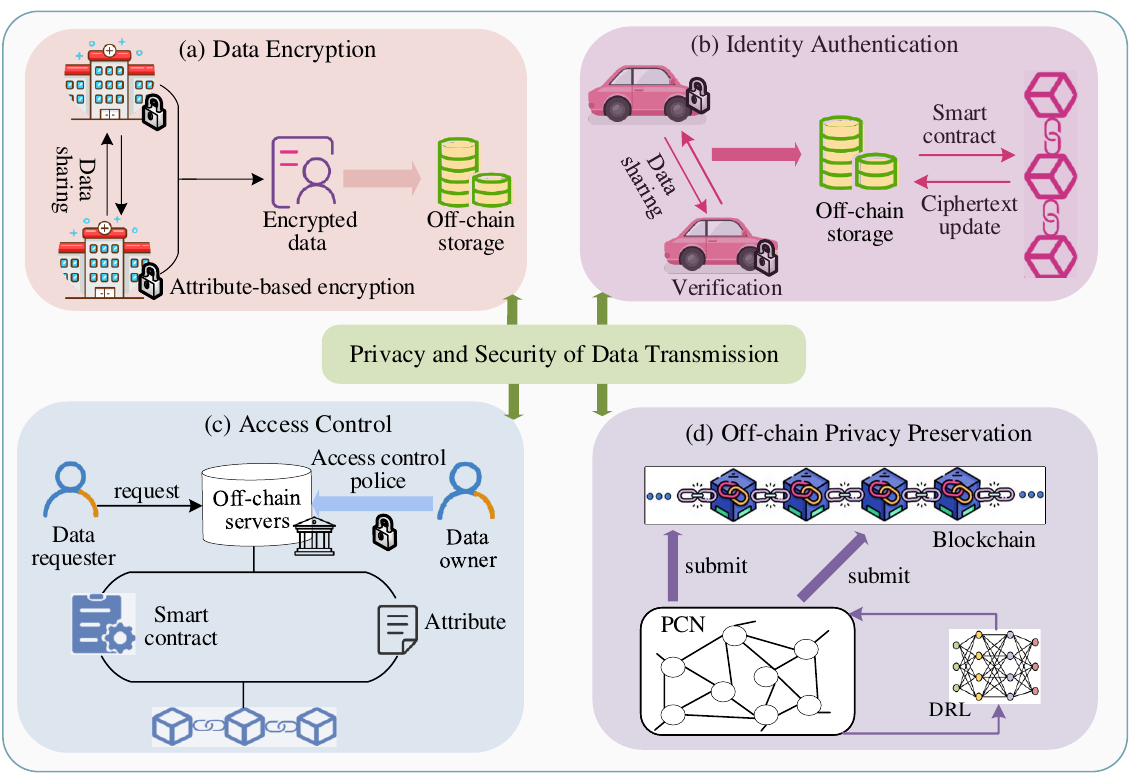}
	\caption{Illustrative solutions of privacy and security of data transmission: (a) Data encryption: hospitals utilize encryption technology to ensure the security of case transmission and prevent it from being maliciously leaked; (b) Identity authentication: vehicles leverage off-chain storage, smart contracts, and blockchain to authenticate users and ensure data security; (c) Access control: users utilize off-chain servers, attribute-based access control, and smart contracts to achieve blockchain access control; (d) Off-chain privacy preservation: PCNs and DRL training models are integrated for off-chain privacy protection.}
	\label{6:Solution.}
	\Description{11}
		\vspace{-0.3cm}
\end{figure}

\indent This subsection focuses on four solutions to privacy and security problems of data transmission, i.e., data encryption, identity authentication, access control and off-chain privacy preservation, as shown in Figure~\ref{6:Solution.}. In the following, we elaborate them in detail.

\subsubsection{Data Encryption}

\indent To improve network security during data transmission, authors in~\cite{Ma2021Research,lin2021model,li2022efficient,Xue2022Blockchain} introduce data encryption technology. It enables sensitive information to be stored in encrypted form off-chain, while allowing interoperability with data on the blockchain when needed, ensuring data privacy and integrity.

\indent Traditional blockchain technology has provided data security in response to the issues of centralization and data silos in the healthcare field. However, it also faces challenges such as excessive data amount and the potential threat of quantum attacks. EMRs typically contain sensitive information about patients and hospitals, making them highly vulnerable to attacks. Authors in~\cite{Ma2021Research} construct a collaborative on-chain and off-chain security system based on symmetric and asymmetric encryption algorithms, realizing the collaborative protection of on-chain and off-chain data. In this system, each device uses its own public and private keys to encrypt and decrypt data, effectively reducing the risk of data tampering during transmission. Similar with~\cite{Ma2021Research}, authors in~\cite{lin2021model} consider storing data off-chain to reduce on-chain storage pressure. The edge node receives data from the off-chain device and performs model training locally, and then the encrypted model parameters are stored by the off-chain storage layer. Different from studies in~\cite{lin2021model} and~\cite{Ma2021Research}, authors in~\cite{li2022efficient} consider a secure keyword searchable attribute-based encryption scheme and introduce the on-chain ledger and off-chain storage model to manage EMR data. The attribute encryption scheme stores the encrypted medical data on off-chain servers with designated storage addresses. Additionally, an index is created and uploaded to the on-chain ledger. Thus, the on-chain ledger and off-chain storage model enables access to medical data, alleviating issues like data silos and limitations in on-chain storage.

\indent Traditional ciphertext updates incur large computational and communication overheads. To this end, authors in~\cite{Xue2022Blockchain} propose a key update mechanism using identity-based hierarchical encryption. This method uses the keyword and time validity to encrypt different types of data. Subsequently, smart contracts are utilized to facilitate the updating of keys and ciphertexts.

\subsubsection{Identity Authentication}

\indent In the blockchain field, identity authentication can be achieved through private and public keys. The former is used to sign transactions and prove the user's identity, while the latter is used to verify the signatures and ensure the legitimacy of transactions. Due to the distributed nature of blockchain and the public nature of transactions, this can cause some security issues if all authentication information is stored on-chain. To address this issue, some researchers have introduced off-chain technologies into blockchain systems, aiming to store transaction and identity verification information off-chain to protect data.

\indent Authors in~\cite{Liu2022Blockchain} propose an anonymous identity verification scheme based on pseudonyms. To ensure secure and reliable off-chain communications, they establish a secure and authenticated communication channel for the sky-air-ground integrated vehicular network, which is not recorded on-chain. To address the identity privacy issues caused by data sharing, authors in~\cite{Xue2022Blockchain} propose a solution for protecting identity privacy in the context of data sharing. ML is used to extract keywords and allows efficient key updates for multiple users in case of key leakage, thus protecting data privacy in an off-chain manner.

\indent Furthermore, the collaboration of different devices on the same task poses significant challenges regarding security and privacy during the communication process. To solve this issue, authors in~\cite{Shen2020Blockchain} propose a solution for secure device authentication. This approach utilizes identity-based signatures and blockchain-assisted secure device authentication to encrypt user's identity without relying on a trusted third party. The user utilizes private keys to generate a signature and the recipient uses public keys to verify it, storing domain-specific information off-chain. Subsequently, entities in different administrative domains authenticate without knowing their true identity, guaranteeing their anonymity and privacy.

\subsubsection{Access Control}

\indent In off-chain networks, access control technology plays a crucial role, which aims to ensure that only authorized users and entities can access specific resources while protecting data privacy and security. Since traditional access control mainly relies on centralized servers, it often faces issues such as single point of failure. Authors in~\cite{Hao2022Smart} link blockchain and attribute-based access control to design a smart contract-based access control framework. Additionally, a set of attributes are considered based on off-chain signatures for distribution, which greatly reduces the storage space on the blockchain and ensures the reliability of off-chain information through on-chain signature verification.

\indent In order to ensure security in the data-sharing process, authors in~\cite{Xue2022Blockchain} propose a blockchain-based access control solution. The approach employs off-chain storage and utilizes various access policies to encrypt different types of data, thereby achieving flexible data access control. Unlike the study in~\cite{Xue2022Blockchain}, authors in~\cite{li2022efficient} combine on-chain and off-chain to design an EMR management model and present a secure keyword-searchable attribute-based encryption solution based on lattice cryptography. In this approach, the actual EMR data of medical institutions is stored on local servers and the corresponding indexes of EMRs are stored on-chain. Therefore, this avoids direct manipulation and malicious tampering  of EMR data.

\subsubsection{Off-chain Privacy Preservation}

\indent It refers to the protection of data processing and storage outside the main blockchain, providing a possible solution to the scalability and privacy challenges faced by blockchain. A scheme to perform secure off-chain cross-channel transfers is proposed in~\cite{Zhang2021Boros}. The opening and closing of channels are publicly traded on the chain, and the detailed information of off-chain transactions is not publicly available, to protect the information of off-chain transactions.

\indent  Generally, most PCNs do not publish the current channel balance. However, transaction senders need to probe the available channel balance of the candidate path, which brings privacy risks. To solve this issue, authors in~\cite{Qiu2022A} propose a privacy-aware transaction scheduling algorithm combining DRL and PCNs. It approximates the transaction scheduling issue as a constrained \textbf{Markov Decision Process (MDP)}, and tries to achieve a balance between transaction throughput and privacy risk. Multi-path payments are considered, which are vulnerable to wormhole attacks. To compensate for this, authors in~\cite{Mazumdar2022Cryptomaze} propose a privacy-preserving atomic multi-path payment protocol that provides privacy protection for user balances. The protocol eliminates the need for multiple off-chain contracts in routing partial payments and aims to improve user privacy and payment security.

\indent To solve the privacy issues in off-chain data transmissions, authors in~\cite{xie2022sofitmix} propose a Bitcoin network mixing protocol based on zero-knowledge proof that can resist \textbf{Denial of Service (DoS)} attacks and support reliable off-chain payments. Additionally, the authors presume that the payee always selects a new address to receive payments in each period. This scheme enhances the reliability of off-chain payments without revealing sensitive information and strengthens user privacy protection. Similar to~\cite{xie2022sofitmix} which uses zero-knowledge proofs to protect privacy, authors in~\cite{Wan2022zk} design a zero-knowledge authentication scheme to protect off-chain data. This scheme follows the strategy of off-chain computing and on-chain verification, providing privacy protection for off-chain data and reducing the computational cost of the blockchain.

\indent \textbf{Lesson 1}: The drawback of encryption technology is its high computational complexity and strict requirements for key updating and revocation, especially obvious in highly dynamic multi-hop routing. Since the complexity of different encryption algorithms is related to their security levels, the trade-off between security and complexity is a challenge. Anonymous identity authentication is achieved through pseudonyms, and it is necessary to further control the communication overhead and improve the identity authentication mechanism. While ensuring off-chain privacy protection, it is also essential to balance the relationship between privacy and performance, and prevent excessive increases in the computing and communication overhead of the off-chain network.

\indent The research on data privacy and security protection mentioned primarily focuses on encryption technology, identity authentication, access control, and off-chain privacy preservation. However, some areas are still overlooked in current research and require further investigation, such as multi-level privacy protection and lightweight data security.

\subsection{Transaction Throughput}

\indent This subsection focuses on three solutions for enhancing transaction throughput: routing decision making, multi-hop payment, and dynamic congestion control.

\subsubsection{Routing Decision Making}

\indent The required number of hops and congestion conditions influence the selection of effective routing. For different paths with the same number of hops, the route with the lowest forwarding cost has priority to be selected since it leads to higher transaction throughput. The studies in~\cite{Mazumdar2022Cryptomaze,Zhang2023Anonymous,Qiu2022A,Malavolt2017Concurrency,xie2022sofitmix} also consider privacy protection while improving transaction throughput. 

\indent Different from payments by one routing path, authors in~\cite{Mazumdar2022Cryptomaze} conduct transaction payments through multiple paths. The proposed algorithms aim to achieve higher throughput and less routing time while protecting privacy. Authors in~\cite{Qiu2022A} use similar multi-path transmission protocol with that in~\cite{sivaraman2020high,Egger2019Atomic}, which achieves high throughput by splitting transactions and sending them over different paths. To balance privacy constraints and transaction throughput, all users of PCNs are assumed to be honest. Then, the authors use Lagrangian and distributed training frameworks for routing optimization to improve the long-term throughput of PCNs. The authors in~\cite{xie2022sofitmix} design a hybrid scheme to resist DoS attacks by using zero-knowledge proofs and Bitcoin, and aim to improve the reliability of off-chain routing decisions and payments.

\subsubsection{Multi-hop Payment}

\indent Previous studies show that PCNs allow two nodes that are not directly connected to conduct payment transactions through a multi-hop path. Participants on the payment path use HTLC and other multi-signature contracts to reach agreements, ensuring secure off-chain transactions. To improve transaction throughput, authors in~\cite{Zhang2023Anonymous,Luo2022Learning,PengW2019Flash,Malavolt2017Concurrency} propose different solutions.

\indent If all transactions are executed on the blockchain, huge overhead and high latency can be introduced. To solve this problem, most research focuses on off-chain multi-hop payments to reduce transaction overheads and latency. Authors in~\cite{Malavolt2017Concurrency} propose a multi-hop solution based on HTLC to enable two nodes to conduct off-chain payments through multi-hop paths. The approach aims to ensure the privacy of transacting parties along the payment path and improve throughput. Similar to~\cite{Malavolt2017Concurrency}, authors in~\cite{Qiu2022A} study the use of HTLC in PCNs to ensure transaction atomicity, authorizing two indirectly connected nodes to conduct off-chain transactions through multi-hop paths. Moreover, in the area of off-chain payment channels, authors in~\cite{Ge2023Magma} extend two-party off-chain payment channels to multi-party off-chain payment channels. This study allows parties to flexibly join in or exit a channel without violating balance security, further enabling instant confirmation and low-cost transactions. 

\indent To avoid transaction failures due to deadline expiration, authors in~\cite{Luo2022Learning} specify transaction priority through forwarding per-hop cost along the transaction path, i.e., high-priority transactions are prioritized for forwarding. The solution assigns different priorities to payment paths through multi-hop priority policies and uses forwarding fees as priority identifiers to balance transaction rates and forwarding costs. Authors in~\cite{Heilman2017Tumblebit,Green2017Bolt,Tairi2021A2L,Zhang2018Anonymous} provide privacy protection with only single-hop payments. To address the privacy problem of PCNs with multi-hop payments, authors in~\cite{Zhang2023Anonymous} focus on anonymous multi-hop payments based on anonymous multi-hop locks. Herein, intermediate users do not know all node information except for neighboring nodes. Additionally, the authors design an anonymous multi-hop payment scheme based on bilinear pairing, which tries to address the high communication overhead caused by anonymous multi-hop locks and increase transaction throughput.

\subsubsection{Dynamic Congestion Control}

\indent Multi-path routing is optimized to improve throughput in PCNs. However, the atomicity of transactions comes at the cost of latency due to network congestion, which affects the delay-sensitive user experience in PCNs. To address this issue, authors in~\cite{Papadis2022Single} propose a single-hop transaction scheduling scheme considering transaction deadlines, but the deadline is uncertain. To this end, authors in~\cite{Wang2022DPCN} propose a deadline-aware PCN system framework in PCNs. They adopt deadline-aware transaction scheduling and transaction congestion avoidance algorithms to perform transactions. Meanwhile, transactions with queuing time exceeding a threshold are marked, which aims to ensure that delay-sensitive transactions can be completed before the deadline, and improve the transaction success rate of PCNs.

\indent To relieve congested channels, authors in~\cite{Zhang2021Robustpay} adopt a multi-path scheduling approach. This method sends transactions through different paths but generates additional load traffic. To solve the severe congestion issues of concurrent PCNs, the study in~\cite{yu2018coinexpress} proposes a fast dynamic routing mechanism that focuses on finding routes with sufficient funds to transmit overloaded transactions. However, the authors only solves partial payment channel congestion issues. Different from the fast routing mechanism in~\cite{yu2018coinexpress}, authors in~\cite{Luo2022Learning} consider priority-aware PCNs. Each intermediate user maintains a priority queue for transactions. The authors propose a priority assignment algorithm based on multi-agent RL, and dynamically estimates the congestion level of each intermediate node based on transaction responses. 

\indent To promote channel balance and maximize throughput simultaneously, authors in~\cite{sivaraman2020high} propose a multi-path transport protocol in PCNs. Under this protocol, each user establishes a router-signaling queue dynamically describing the real-time congestion size of the channel. When the transaction is paid unidirectionally for a long time, leading to the depletion of channel balance in that direction, it results in congestion scenarios. Thus, the protocol sends congestion signals to suppress transactions in that direction. 

\indent \textbf{Lesson 2}: In congested networks, transaction latency becomes high. Existing research mostly assumes that users are honest, which is an ideal situation. How to solve the problem caused by malicious users in off-chain transactions while ensuring high throughput is a research direction worth exploring.

\indent Different off-chain transaction rate may have different transaction demands. In off-chain payments, existing solutions typically consider the impact of forwarding fees, deadlines, transaction congestion, privacy protection, and other factors on multi-hop payments separately, but do not comprehensively consider the impact of these factors at each individual hop. In addition, multi-hop payments are vulnerable to various attacks that may compromise the security and privacy of transactions. However, there is little research on the security and privacy protection of multi-hop payments.

\subsection{Data Storage and Sharing}

\indent This subsection focuses on two solutions for data storage and sharing: off-chain storage and off-chain-based smart contract.

\subsubsection{Off-chain Storage}

\begin{figure}
	\centering
	\includegraphics[scale = 0.35]{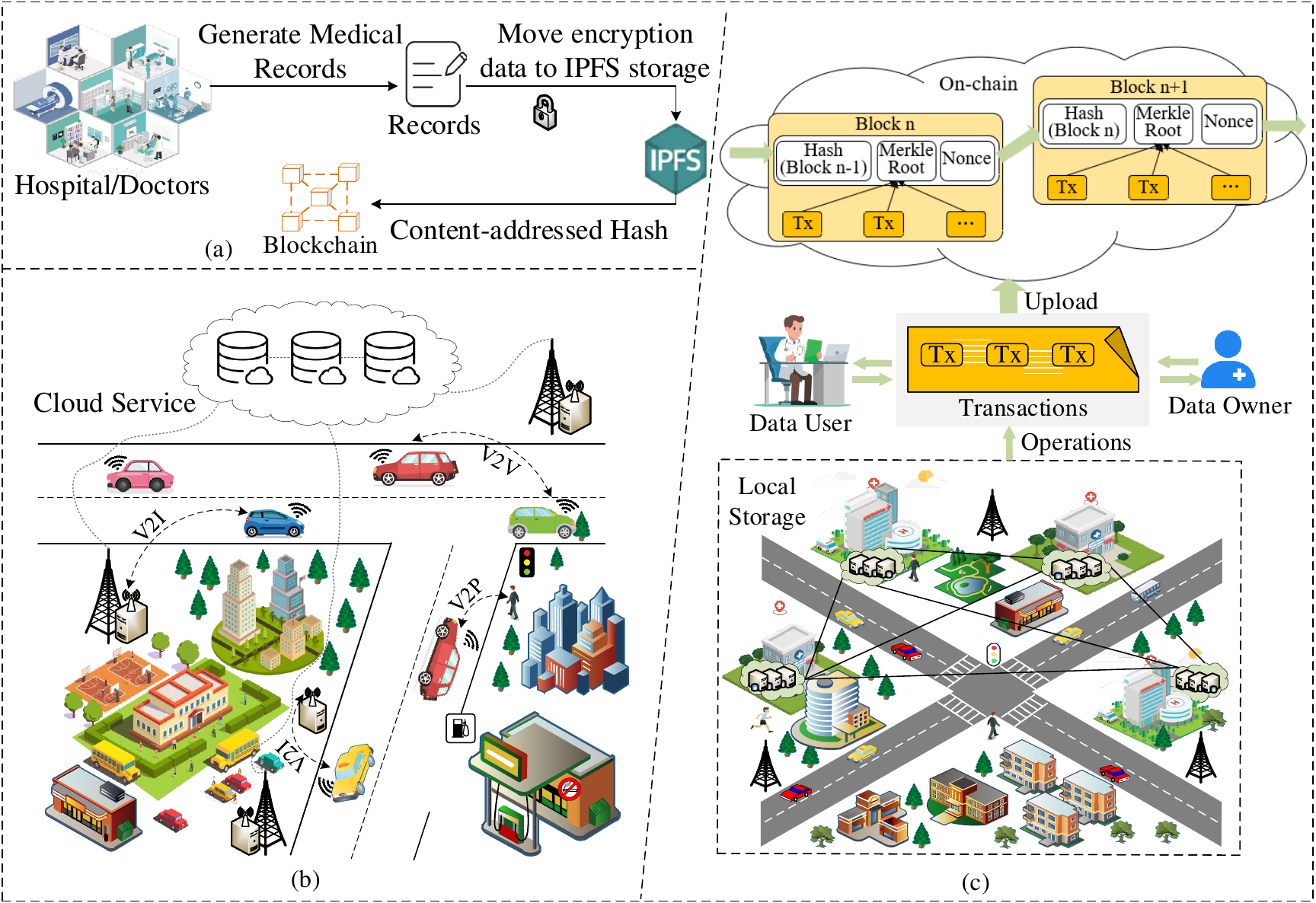}
	\caption{Off-chain storage: (a) IPFS storage, (b) Cloud storage, and (c) Local storage.}
	\label{7:S3.}
	\Description{11}
		\vspace{-0.3cm}
\end{figure}

\indent Due to the storage limitations of blockchain, many researchers have considered off-chain storage as a viable solution to enhance blockchain scalability, such as IPFS~\cite{jayabalan2022scalable,Azbeg2022Access,wang2021lifesaving,Liang2022PDPChain}, cloud storage~\cite{Shen2020Blockchain,lin2021model,Xue2022Blockchain}, local storage~\cite{li2022efficient} as shown in Figure~\ref{7:S3.}.

\indent Considering the limited battery and computing capacity of UAVs, they may not be able to handle heavy emergency tasks. Authors in~\cite{wang2021lifesaving} propose a safe and efficient information sharing solution for UAV disaster relief. Specifically, this method transfers the heavy storage and processing tasks from resource-constrained UAVs to off-chain ground vehicles. The authors use IPFS as an off-chain data repository to record extensive source data, and on-chain records data pointers.

\indent Currently, hospitals commonly adopt electronic methods to record patient cases and treatment plans, with the majority of EMRs stored in a centralized manner. However, centralized storage is difficult to protect patient privacy effectively, and poses security risks to patients' sensitive information. To address this issue, most researchers integrate blockchain with IPFS for off-chain medical data storage~\cite{jayabalan2022scalable,Azbeg2022Access}. The study in~\cite{Azbeg2022Access} stores medical data in IPFS, and combines an Ethereum blockchain based on proof-of-authority to accelerate data storage. In addition, the authors incorporate an encryption mechanism to ensure data security. Similar to~\cite{Azbeg2022Access}, authors in~\cite{jayabalan2022scalable} use IPFS for off-chain storage to manage EMRs efficiently. They protect patient data privacy through access control, distributed storage, and encryption. To meet the extensive storage requirements of medical data, the authors store patients' encrypted data in the IPFS system to provide an additional layer of security. IPFS assigns a unique hash address to each file uploaded to the system, enabling content addressing. Only IPFS addresses are stored to reduce the transaction size. The study in~\cite{Liang2022PDPChain} presents a personal data privacy protection scheme, which leverages an improved Paillier homomorphic encryption mechanism to encrypt the original data, and then upload it to IPFS to obtain the hash value. Additionally, the authors combine blockchain and distributed private clusters for off-chain storage of encrypted data.

\indent Different from leveraging IPFS as off-chain data storage, the study in~\cite{li2022efficient} uses local servers as off-chain storage. Authors store encrypted EMR data on local servers and create an index to record it on the blockchain, thereby alleviating the heavy burden on the blockchain. The solution ensures the security of EMRs, and also offers convenience to users seeking treatment at different healthcare institutions. Different from~\cite{li2022efficient}, authors in~\cite{Shen2020Blockchain,lin2021model,Xue2022Blockchain} use cloud servers for data storage. Authors in~\cite{lin2021model} store the parameters of the locally trained model in the off-chain storage layer consisting of cloud servers, edge servers and devices. Authors in~\cite{Xue2022Blockchain} store the generated ciphertext on cloud servers to reduce data storage and management costs. After the requester pays, the data owner provides the decryption key to the requester for data sharing.

\subsubsection{Off-chain-based Smart Contract} 

\indent Smart contracts are a computer protocol designed to digitally propagate, validate, and execute contracts, avoiding the third-party oversight in traditional transaction systems, since it is independent of the third-party to build a trustworthy environment~\cite{Yue2021A}. In addition, off-chain contracts are a type of smart contract, where the logic encoded in the contract is not executed by miners, but by the participants who instantiate the contract. The advantage is that computationally intensive tasks can be performed off-chain as long as the participants behave honestly~\cite{Mazumdar2022Cryptomaze}.

\indent With the emergence of smart contracts, their applications have expanded to data storage and sharing. The studies in~\cite{Xue2022Blockchain,Chen2020Toward} link smart contracts with data sharing, both of which use thr cloud to store data. Authors in~\cite{Xue2022Blockchain} design a smart blockchain-based data-sharing scheme, which generates update tokens through smart contracts to promote key and ciphertext updates, and stores data on cloud servers. Authors in~\cite{lin2021model} propose a model training mechanism based on on-chain and off-chain collaboration. In order to prevent edge training nodes from intentionally uploading wrong parameters, the authors design a reputation evaluation mechanism, which uses the credibility factor smart contract to dynamically adapt the reputation of edge nodes and improve the aggregation quality of the model. To solve the data privacy and authenticity issues of smart contracts, authors in~\cite{Wan2022zk} design an off-chain data feed scheme with zero-knowledge authentication, which ensures the authenticity and privacy of off-chain data transmitted to smart contracts.

\indent Oracles are a core mechanism for obtaining off-chain data on the blockchain, providing data request services for other contracts. When smart contracts on the blockchain have data interaction needs, the oracle helps them to collect data off-chain and upload the acquired data to the on-chain smart contracts after validation. Hence, some researchers utilize this technology to facilitate the connection between the blockchain and the off-chain world~\cite{pasdar2023connect}.

\indent \textbf{Lesson 3}: We notice that current studies mainly focus on IPFS, cloud, and local storage to reduce the storage pressure on-chain and expand the performance of the blockchain. In the design of off-chain smart contracts, the security and privacy protection of off-chain networks, as well as how to efficiently submit off-chain results to the blockchain should be comprehensively considered. Thus, factors such as efficiency, cost, and reliability bring new research challenges for off-chain data storage and sharing. Another challenge is how to adopt lightweight off-chain storage for large volumes of transactions. It is crucial to address the issue of data synchronization and updating, and identify appropriate strategies to ensure data consistency and reliability. What's more, how to well coordinate on-chain and off-chain data is also worthy of further investigation.

\subsection{Trusted Off-chain Data}

\indent This subsection focuses on two solutions for trusted off-chain data: trust evaluation and computing, and incentive mechanism.

\subsubsection{Trust Evaluation and Computing}

\indent Off-chain nodes are not the main component of core blockchain networks but perform various tasks, such as transactions processing, and data storing. To ensure that these off-chain nodes perform tasks honestly and efficiently, a trust-based computing mechanism needs to be established. Existing blockchain technologies have been used in vehicular data sharing~\cite{Chen2019A}. However, it is not efficient in the UAV-assisted IoV for disaster relief, due to its heavy reliance on the availability of RSUs and the PoW consensus algorithm.

\indent If the network infrastructure is damaged in post-disaster areas, the network performance may be adversely affected by interruptions, increased latency, and communication barriers. Authors in~\cite{Su2022LVBS} establish a credit assessment mechanism in the vehicular blockchain environment to track misbehavior and reward the honest behavior of legitimate nodes. This mechanism can be applied between off-chain ground vehicles and UAVs, assisting in disaster relief efforts by permanently recording and tracking behavior records. However, when vehicles share data with RSUs and other vehicles, the data provided by vehicles is not always trustworthy. In addition, a large amount of data sharing may cause channel congestion. To solve the above issues, authors in~\cite{Yuan2023TRUCON} propose a blockchain-based trusted data-sharing mechanism with congestion state. The authors control the channel congestion state using the Kademlia algorithm-based data forwarding mechanism for connected vehicular traffic, to limit the number of forwarding vehicles and reduce the channel communication overhead. By combining both on-chain and off-chain components, a trust management system with congestion control is designed to ensure data accuracy. 

\indent To improve the quality of data provided by the physical world off-chain, authors in~\cite{lin2021model} design a reputation evaluation mechanism based on confidence factors to avoid uploading erroneous data. Based on the feedback results of reputation and confidence factors, this solution only aggregates data with high reputation and high confidence factors to ensure data accuracy. To address the lack of transparency faced by traditional access control, authors in~\cite{Hao2022Smart} propose a smart contract-based access control framework, enabling trustworthy access decisions for clients. The owner assigns each client a set of attributes, and the client's address is contained in the off-chain signature signed by the owner, which is evaluated on-chain to make trustworthy access decisions. This solution aims to the reliability of off-chain attribute data through efficient on-chain signature verification.

\indent Since companies continue connecting their manufacturing and logistics infrastructures to the IoT, trustworthiness and privacy issues are becoming increasingly important. Current research has proposed blockchain as a solution, but its scalability is limited and may compromise user privacy. The authors in~\cite{zhang2020off} propose off-chain trusted computing to ensure data authenticity and privacy. They offload massive transaction workload to TEEs established by off-chain computing nodes, while leaving only the execution of business logic to on-chain computing to improve transaction efficiency and speeds. Off-chain trusted computing is a hybrid computing paradigm, where each trusted computing node has a pair of asymmetric keys, and each dataset is encrypted with a randomly generated symmetric key to protect information privacy. The selected worker performs trusted computing tasks, and the data encryption key is protected and the worker's asymmetric key is used distributively. Authors in~\cite{Weng2022Golden} design a secure off-chain and on-chain interaction protocol based on TEE to ensure the integrity and authenticity of the benchmark tests. Additionally, the authors ensure that the model submitted within TEE is the same as the model submitted on the blockchain, so that the on-chain commitment to store the model off-chain is trustworthy.

\begin{table*}
	\caption{Summary of solutions for off-chain networks}
	\label{table3}
	
	\centering
	\linespread{1.1}\selectfont
	\renewcommand\arraystretch{1.05}
	\begin{tabular}{|*{12}{c|}}
		\hline
		\multirow{10}*{\parbox[t]{4mm}{\centering\textbf{Ref.}}} & 
		\multirow{10}*{\parbox[t]{32mm}{\centering\textbf{Description}}}&
		\multirow{10}*{\parbox[t]{14mm}{\centering\textbf{Technolo- gies}}}&
		\multirow{10}*{\parbox[t]{14mm}{\centering\textbf{Storage methods}}}&
		\multirow{10}*{\parbox[t]{16mm}{\centering\textbf{Applicati-\\on scenarios}}}&
		\multicolumn{7}{c|}{\parbox[t]{25mm}{\centering\textbf{Optimization objectives}}}  \\
		
		\cline{6-12}
		&&&&&
		
		\parbox[t]{2mm}{\multirow{8}*{\rotatebox[origin=c]{90}{\vspace{28mm}\textbf{Privacy and security}}}} &
		\parbox[t]{2mm}{\multirow{8}*{\rotatebox[origin=c]{90}{\textbf{Delay}}}} & 
		\parbox[t]{2mm}{\multirow{8}*{\rotatebox[origin=c]{90}{\textbf{Throughput}}}} & 
		\parbox[t]{2mm}{\multirow{8}*{\rotatebox[origin=c]{90}{\textbf{Costs}}}} & 
		\parbox[t]{2mm}{\multirow{8}*{\rotatebox[origin=c]{90}{\textbf{Accuracy}}}} & 
		\parbox[t]{2mm}{\centering\multirow{8}*{\rotatebox[origin=c]{90}{\textbf{Transaction success rate}}}} &
		\parbox[t]{2mm}{\centering\multirow{8}*{\rotatebox[origin=c]{90}{\textbf{Efficiency}}}} \\
		&&&&&&&&&&& \\
		&&&&&&&&&&& \\
		&&&&&&&&&&& \\
		&&&&&&&&&&& \\
		&&&&&&&&&&& \\
		&&&&&&&&&&& \\
		&&&&&&&&&&& \\
		\hline

		\multirow{4}*{\cite{lin2021model}}&	
		\multirow{4}{1.5in}{A model training mechan-\\ism based on on-chain and off-chain collaboration for edge computing}&	
		\multirow{4}{0.6in}{\centering{Edge com-\\puting; Data collaboration} }	&	
		\multirow{4}{0.6in}{\centering{Off-chain storage}}	&
		\multirow{4}{0.6in}{\centering{Blockchain}}&	
		\multirow{4}*{$\times$}	&
		\multirow{4}*{$\times$}&		
		\multirow{4}*{$\times$}&
		\multirow{4}*{$\times$}	&
		\multirow{4}*{$\surd$}&
		\multirow{4}*{$\times$}	&
		\multirow{4}*{$\surd$}	

		\\ &&&&&&&&&&&  \\  &&&&&&&&&&& \\ &&&&&&&&&&&  \\ \cline{1-12}

		\multirow{2}*{\cite{Weng2022Golden}}&	
		\multirow{2}{1.5in}{A secure off-chain and on-\\chain interaction protocol   }&  
		\multirow{2}{0.6in}{\centering{Blockchain; TEE}}	&	
		\multirow{2}{0.6in}{\centering{Off-chain storage}}	&
		\multirow{2}{0.6in}{\centering{Blockchain}}	&	
		\multirow{2}*{$\surd$}	&
		\multirow{2}*{$\times$}&	
		\multirow{2}*{$\times$}&	
		\multirow{2}*{$\times$}&
		\multirow{2}*{$\times$}&
		\multirow{2}*{$\times$}&
		\multirow{2}*{$\times$}  
		
		\\	&&&&&&&&&&& \\   \cline{1-12}

			\multirow{3}*{\cite{wang2021lifesaving}}&	
		\multirow{3}{1.5in}{A secure off-chain infor-\\mation sharing scheme for UAV-aided disaster rescue  }&
		\multirow{3}{0.6in}{\centering{RL; Vehi-\\cular fog computing}}	&	
		\multirow{3}{0.55in}{\centering{Off-chain storage}}	&
		\multirow{3}{0.6in}{\centering{Disaster rescue}}	&	
		\multirow{3}*{$\times$}	&
		\multirow{3}*{$\times$}&	
		\multirow{3}*{$\times$}&	
		\multirow{3}*{$\surd$}&
		\multirow{3}*{$\times$}&
		\multirow{3}*{$\times$}&
		\multirow{3}*{$\times$}        
		
		\\	&&&&&&&&&&& \\ &&&&&&&&&&& \\   \cline{1-12}

		\multirow{4}*{\cite{Liu2022Blockchain}}&	
		\multirow{4}{1.5in}{A blockchain-based collab-\\orative credential manage-\\ment scheme for anony-\\mous authentication }&
		\multirow{4}{0.6in}{\centering{Zero-knowledge proof}}	&	
		\multirow{4}{0.6in}{\centering{Off-chain storage}}	&
		\multirow{4}{0.6in}{\centering{Vehicular network}}	&	
		\multirow{4}*{$\surd$}	&
		\multirow{4}*{$\times$}&	
		\multirow{4}*{$\times$}&	
		\multirow{4}*{$\times$}&
		\multirow{4}*{$\times$}&
		\multirow{4}*{$\times$}&
		\multirow{4}*{$\surd$}
		\\ &&&&&&&&&&&  \\  &&&&&&&&&&& \\ &&&&&&&&&&& \\ \cline{1-12}
		
		\multirow{4}*{\cite{Qiu2022A}}&	
		\multirow{4}{1.5in}{A privacy-aware routing algorithm with defence against DoS attacks based on DRL  }&
		\multirow{4}{0.6in}{\centering{DRL}}	&	
		\multirow{4}{0.6in}{\centering{Off-chain storage}}	&
		\multirow{4}{0.6in}{\centering{Off-chain transaction}}	&	
		\multirow{4}*{$\surd$}	&
		\multirow{4}*{$\times$}&	
		\multirow{4}*{$\surd$}&	
		\multirow{4}*{$\times$}&
		\multirow{4}*{$\times$}&
		\multirow{4}*{$\times$}&
		\multirow{4}*{$\times$}
		\\ &&&&&&&&&&&  \\  &&&&&&&&&&& \\    &&&&&&&&&&& \\  \cline{1-12}

				\multirow{4}*{\cite{Shen2020Blockchain}}&	
		\multirow{4}{1.5in}{A blockchain-assisted secure device authentica-\\tion mechanism }&
		\multirow{4}{0.65in}{\centering{Identity-based sign-\\ature technologies}  }	&	
		\multirow{4}{0.55in}{\centering{Cloud server}}	&
		\multirow{4}{0.6in}{\centering{Blockchain}}	&	
		\multirow{4}*{$\surd$}	&
		\multirow{4}*{$\times$}&	
		\multirow{4}*{$\times$}&	
		\multirow{4}*{$\times$}&
		\multirow{4}*{$\times$}&	
		\multirow{4}*{$\times$}&
		\multirow{4}*{$\times$}
		\\	&&&&&&&&&&& \\  &&&&&&&&&&& \\   &&&&&&&&&&& \\ \cline{1-12}

		\multirow{3}*{\cite{Wan2022zk}}&	
		\multirow{3}{1.5in}{A zero-knowledge authen-\\ticated off-chain data feed scheme }&
		\multirow{3}{0.6in}{\centering{Zero-knowledge proof}}	&	
		\multirow{3}{0.6in}{\centering{Off-chain storage}}	&
		\multirow{3}{0.6in}{\centering{Blockchain}}	&	
		\multirow{3}*{$\surd$}	&
		\multirow{3}*{$\times$}&	
		\multirow{3}*{$\times$}&	
		\multirow{3}*{$\times$}&
		\multirow{3}*{$\surd$}&
		\multirow{3}*{$\times$}&
		\multirow{3}*{$\times$}	
		\\ &&&&&&&&&&&  \\  &&&&&&&&&&& \\  \cline{1-12}
		
				\multirow{3}*{\cite{Zhang2021Boros}}&	
		\multirow{3}{1.5in}{An off-chain channel hub system  }&
		\multirow{3}{0.6in}{\centering{Channel hub protocol}}	&	
		\multirow{3}{0.6in}{\centering{N/A}}	&
		\multirow{3}{0.6in}{\centering{Off-chain transaction}}	&	
		\multirow{3}*{$\surd$}	&
		\multirow{3}*{$\times$}&	
		\multirow{3}*{$\times$}&	
		\multirow{3}*{$\times$}&
		\multirow{3}*{$\times$}&
		\multirow{3}*{$\times$}&
		\multirow{3}*{$\times$}
		\\	&&&&&&&&&&& \\  	&&&&&&&&&&& \\   \cline{1-12}

				\multirow{3}*{\cite{Wang2022DPCN}}&	
		\multirow{3}{1.5in}{A deadline-aware PCN framework  }&
		\multirow{3}{0.6in}{\centering{Congestion avoidance algorithm}}	&	 
		\multirow{3}{0.55in}{\centering{Off-chain storage}}	&
		\multirow{3}{0.6in}{\centering{PCNs}}	&	
		\multirow{3}*{$\times$}	&
		\multirow{3}*{$\surd$}&	
		\multirow{3}*{$\times$}&	
		\multirow{3}*{$\times$}&
		\multirow{3}*{$\times$}&
		\multirow{3}*{$\surd$}&
		\multirow{3}*{$\times$}
		\\	&&&&&&&&&&& \\   &&&&&&&&&&& \\ \cline{1-12}

		\hline
		\multicolumn{12}{c}{(``$\surd$'' if the solution satisfies the property, ``$\times$'' if not)}
	\end{tabular}

\end{table*}

\begin{table*}
	\caption{Summary of solutions for off-chain networks (Cont.)}
	\label{table4}
	
	\centering
	\linespread{1.1}\selectfont
	\renewcommand\arraystretch{1.05}
	\begin{tabular}{|*{12}{c|}}
		\hline
		\multirow{10}*{\parbox[t]{4mm}{\centering\textbf{Ref.}}} & 
		\multirow{10}*{\parbox[t]{32mm}{\centering\textbf{Description}}}&
		\multirow{10}*{\parbox[t]{14mm}{\centering\textbf{Technolo- gies}}}&
		\multirow{10}*{\parbox[t]{14mm}{\centering\textbf{Storage methods}}}&
		\multirow{10}*{\parbox[t]{16mm}{\centering\textbf{Applicati-\\on scenarios}}}&
		\multicolumn{7}{c|}{\parbox[t]{25mm}{\centering\textbf{Optimization objectives}}}  \\
		
		\cline{6-12}
		&&&&&
		
		\parbox[t]{2mm}{\multirow{8}*{\rotatebox[origin=c]{90}{\vspace{25mm}\textbf{Privacy and security}}}} &
		\parbox[t]{2mm}{\multirow{8}*{\rotatebox[origin=c]{90}{\textbf{Delay}}}} & 
		\parbox[t]{2mm}{\multirow{8}*{\rotatebox[origin=c]{90}{\textbf{Throughput}}}} & 
		\parbox[t]{2mm}{\multirow{8}*{\rotatebox[origin=c]{90}{\textbf{Costs}}}} & 
		\parbox[t]{2mm}{\multirow{8}*{\rotatebox[origin=c]{90}{\textbf{Accuracy}}}} & 
		\parbox[t]{2mm}{\centering\multirow{8}*{\rotatebox[origin=c]{90}{\textbf{Transaction success rate}}}} &
		\parbox[t]{2mm}{\centering\multirow{8}*{\rotatebox[origin=c]{90}{\textbf{Efficiency}}}} \\
		&&&&&&&&&&& \\
		&&&&&&&&&&& \\
		&&&&&&&&&&& \\
		&&&&&&&&&&& \\
		&&&&&&&&&&& \\
		&&&&&&&&&&& \\
		&&&&&&&&&&& \\
		\hline

		\multirow{5}*{ \cite{Luo2022Learning}}&	
		\multirow{5}{1.5in}{A priority-aware transac-\\tion forwarding mechani-\\sm and a multi-agent DQN-based priority assig-\\nment algorithm}&
		\multirow{5}{0.6in}{\centering{Multi-agent RL; Priority scheduling}}	&	
		\multirow{5}{0.55in}{\centering{N/A}}	&
		\multirow{5}{0.6in}{\centering{Off-chain transaction}}	&	
		\multirow{5}*{$\times$}	&
		\multirow{5}*{$\times$}&	
		\multirow{5}*{$\surd$}&	
		\multirow{5}*{$\times$}&
		\multirow{5}*{$\times$}&
		\multirow{5}*{$\surd$}&	
		\multirow{5}*{$\times$}
		\\ &&&&&&&&&&&  \\  &&&&&&&&&&& \\  &&&&&&&&&&& \\  &&&&&&&&&&& \\ \cline{1-12}

		\multirow{3}*{\cite{Mazumdar2022Cryptomaze}}&	
		\multirow{3}{1.5in}{A secure and privacy-\\preserving atomic multi-\\path payment protocol }&
		\multirow{3}{0.65in}{\centering{Multi-path payment protocol}}	&	
		\multirow{3}{0.6in}{\centering{Off-chain storage}}	&
		\multirow{3}{0.6in}{\centering{Off-chain transaction}}	&	
		\multirow{3}*{$\surd$}	&
		\multirow{3}*{$\surd$}&	
		\multirow{3}*{$\times$}&	
		\multirow{3}*{$\times$}&
		\multirow{3}*{$\times$}&
		\multirow{3}*{$\times$}&
		\multirow{3}*{$\times$}
		\\ &&&&&&&&&&&  \\  &&&&&&&&&&& \\  \cline{1-12}

		\multirow{3}*{\cite{Zhang2021Robustpay}}&	
		\multirow{3}{1.5in}{A distributed robust payment routing protocol against transaction failures  }&
		\multirow{3}{0.65in}{\centering{HTLC; Multi-path scheduling}}	&	
		\multirow{3}{0.55in}{\centering{N/A}}	&
		\multirow{3}{0.6in}{\centering{Off-chain transaction}}	&	
		\multirow{3}*{$\times$}	&
		\multirow{3}*{$\times$}&	
		\multirow{3}*{$\times$}&	
		\multirow{3}*{$\times$}&
		\multirow{3}*{$\times$}&
		\multirow{3}*{$\times$}&
		\multirow{3}*{$\surd$}		
		\\	&&&&&&&&&&& \\ &&&&&&&&&&& \\   \cline{1-12}	
		
		\multirow{3}*{\cite{sivaraman2020high}}&	
		\multirow{3}{1.5in}{A multi-path congestion control protocol }&
		\multirow{3}{0.65in}{\centering{Multi-path congestion control}}	&	
		\multirow{3}{0.55in}{\centering{Off-chain storage}}	&
		\multirow{3}{0.6in}{\centering{PCNs}}	&	
		\multirow{3}*{$\times$}	&
		\multirow{3}*{$\times$}&	
		\multirow{3}*{$\surd$}&	
		\multirow{3}*{$\times$}&
		\multirow{3}*{$\times$}&
		\multirow{3}*{$\times$}&
		\multirow{3}*{$\surd$}
		\\	&&&&&&&&&&& \\  &&&&&&&&&&& \\  \cline{1-12}

				\multirow{3}*{\cite{li2022efficient}}&	
		\multirow{3}{1.5in}{A secure keyword searcha-\\ble attribute-based encryp-\\tion scheme}&	
		\multirow{3}{0.6in}{\centering{Encryption techniques}}	&	
		\multirow{3}{0.6in}{\centering{Local server}}	&
		\multirow{3}{0.6in}{\centering{Healthcare}}&	
		\multirow{3}*{$\surd$}	&
		\multirow{3}*{$\times$}&		
		\multirow{3}*{$\times$}&
		\multirow{3}*{$\times$}	&
		\multirow{3}*{$\times$}&
		\multirow{3}*{$\times$}	&
		\multirow{3}*{$\times$}	
		
		\\ &&&&&&&&&&&  \\  &&&&&&&&&&& \\  \cline{1-12}

				\multirow{3}*{\cite{Xue2022Blockchain}}&	
		\multirow{3}{1.5in}{An intelligent blockchain-\\based data-sharing scheme for future networks }&
		\multirow{3}{0.6in}{\centering{ML; \\Encryption algorithms}}	&	
		\multirow{3}{0.6in}{\centering{Cloud server}}	&
		\multirow{3}{0.6in}{\centering{Data sharing}}	&	
		\multirow{3}*{$\surd$}	&
		\multirow{3}*{$\times$}&	
		\multirow{3}*{$\times$}&	
		\multirow{3}*{$\surd$}&
		\multirow{3}*{$\times$}&
		\multirow{3}*{$\times$}&
		\multirow{3}*{$\times$}
		\\ &&&&&&&&&&&  \\  &&&&&&&&&&& \\  \cline{1-12}

				\multirow{3}*{\cite{Yuan2023TRUCON}}&	
		\multirow{3}{1.5in}{A blockchain-based trust-\\ed data sharing mechanis-\\m with congestion control }&
		\multirow{3}{0.6in}{\centering{Blockchain; Trust management} }	&	
		\multirow{3}{0.55in}{\centering{Vehicle storage}}	&
		\multirow{3}{0.6in}{\centering{IoV}}	&	
		\multirow{3}*{$\times$}	&
		\multirow{3}*{$\times$}&	
		\multirow{3}*{$\times$}&	
		\multirow{3}*{$\times$}&
		\multirow{3}*{$\surd$}&
		\multirow{3}*{$\times$}&	
		\multirow{3}*{$\times$}
		\\	&&&&&&&&&&& \\  &&&&&&&&&&& \\  \cline{1-12}

		\multirow{3}*{\cite{Chen2020Toward}}&	
		\multirow{3}{1.5in}{An auction-based incen-\\tive mechanism based on a consortium blockchain  }&
		\multirow{3}{0.6in}{\centering{Blockchain;\\Incentive mechanism}}	&	
		\multirow{3}{0.55in}{\centering{Off-chian storage}}	&
		\multirow{3}{0.6in}{\centering{IoV}}	&	
		\multirow{3}*{$\surd$}	&
		\multirow{3}*{$\times$}&	
		\multirow{3}*{$\times$}&	
		\multirow{3}*{$\times$}&
		\multirow{3}*{$\surd$}&
		\multirow{3}*{$\times$}&
		\multirow{3}*{$\times$}
		\\	&&&&&&&&&&& \\   &&&&&&&&&&& \\  \cline{1-12}

		\multirow{4}*{\cite{Su2022LVBS}}&	
		\multirow{4}{1.5in}{A UAV and blockchain-\\assisted collaborative air-\\ground network architec-\\ture in disaster areas }&
		\multirow{4}{0.6in}{\centering{Blockchain; RL}  }	&	
		\multirow{4}{0.55in}{\centering{Vehicle storage}}	&
		\multirow{4}{0.6in}{\centering{IoV; Data sharing}}	&	
		\multirow{4}*{$\surd$}	&
		\multirow{4}*{$\times$}&	
		\multirow{4}*{$\times$}&	
		\multirow{4}*{$\times$}&
		\multirow{4}*{$\surd$}&
		\multirow{4}*{$\times$}&
		\multirow{4}*{$\times$}
		\\	&&&&&&&&&&& \\ &&&&&&&&&&& \\ &&&&&&&&&&& \\  \cline{1-12}

		\hline
		\multicolumn{12}{c}{(``$\surd$'' if the solution satisfies the property, ``$\times$'' if not)}
	\end{tabular}
\end{table*}

\subsubsection{Incentive Mechanism}

\indent It refers to the use of various methods to encourage efficient collaborations among multiple users to maximize their interests, and its incentive process can be viewed as a reward transaction process to promote honest user engagement~\cite{Chen2022Timeliness}. In off-chain transactions, intermediate nodes are responsible for forwarding and validating transactions. In the transaction process, with various network attacks, the privacy and security of data are threatened, and users are reluctant to provide data, fearing data leakage and malicious use. For example, in the IoV, users are often reluctant to share data or forward messages, fearing the risk of privacy leakage of sensitive information such as locations.

\indent To encourage the contribution of off-chain ground vehicles and provide high quality data, authors in~\cite{Chen2019CVCG} propose a coalition game-based cooperative vehicle-to-vehicle content dissemination mechanism. However, accurate parameters of the network model for shared services may not be suitable for data providers and consumers. Based on the abovementioned issues, authors in~\cite{Su2022LVBS} propose an RL-based incentive mechanism to encourage high-quality data sharing among ground vehicles by describing the interaction between data providers and consumers as a finite MDP. Similar to~\cite{Chen2019CVCG} and~\cite{Su2022LVBS}, authors in~\cite{Chen2020Toward} propose a quality-driven auction-based incentive mechanism that aims to incentive users to provide high-quality data and uses the expectation-maximization algorithm to evaluate data quality, ensuring the trustworthiness of off-chain data. The authors in~\cite{Gao2018Truthful} design an incentive mechanism considering the data quality of off-chain ground vehicles. It aims to stimulate vehicles to provide truthful data to maximize social welfare.

\indent The incentive mechanism also promotes honest behavior of off-chain nodes and contributes real data. The authors in~\cite{Zhao2022ATensor} design an incentive mechanism based on tensor computing that motivates off-chain ground nodes to provide reliable data and address security issues to maximize social welfare. Currently, researchers have adopted incentive mechanisms to motivate off-chain computation and ensure data trustworthiness. By incentivizing participants to contribute their computing resources to execute computations off-chain and receive corresponding rewards, it can improve the speed of transaction processing and reduce the cost of executing smart contracts. 

\indent \textbf{Lesson 4}: Building a trustworthy environment is of great importance to off-chain networks. The above research primarily focuses on trustworthy off-chain data through trust evaluation, trusted computing and incentive mechanisms. These schemes can be further integrated with encryption techniques to establish a usable trust model while protecting user privacy. However, there are still some issues to be further investigated to achieve trustworthy off-chain networks. For example, in order to establish a trustworthy off-chain network, it is important to gather and analyze as much data as possible. Off-chain networks can leverage tiny ML for model training to achieve low consumption, privacy protection, low latency, and other effects.

\indent Although trust evaluation and trusted computing ensure the trustworthiness of off-chain data, the required computing and storage resources cannot be overlooked. Meanwhile, incentive mechanisms also face issues of unfairness and unreasonable among participants. Last, designing a suitable algorithm for calculating node trust values and enabling efficient off-chain payments presents certain challenges.

\section{Research Challenges and Open Issues}\label{five}

\indent In the previous sections, we review the background, issues, technologies and solutions of off-chain networks. However, some research challenges and open issues still exist, which will be discussed in this section.

\subsection{Security and Privacy Protection of Off-chain Networks}

\indent Different from traditional blockchain networks, off-chain networks process transactions and data outside the main chain, which can improve transaction speeds, reduce transaction costs, increase scalability, and enhance privacy protection. Nevertheless, security and privacy protection remain key challenges for off-chain networks. In off-chain networks, multiple transaction parties usually cooperate to complete transactions. How to manage the privacy of these participants is rather challenge. Furthermore, it is important to consider node security to ensure the overall security of the payment network.

\indent Patient-center privacy protection is an application area of off-chain networks. In order to avoid privacy leakage, some patients with sensitive information are unwilling to share their medical records among medical institutions, resulting in a serious data island phenomenon. The existing privacy protection measures for users' personal information and transaction data, such as anonymity, obfuscation, and encryption techniques, still have issues with complexity and high costs. In addition, authors in~\cite{Green2017Bolt} and~\cite{Heilman2017Tumblebit} propose privacy protection protocols, where all users perform off-chain payments through a unique intermediary or hub. However, it is not clear how these solutions can be extended to multi-hop PCNs. Whether for single-hop or multi-hop payments, most solutions that prevent payment channels from becoming congested do not incorporate user privacy protection. Off-chain payment channels are susceptible to various attacks and still face vulnerabilities and privacy risks. 

\subsection{Transaction Deadline of Payment Channel Networks}

\indent In PCNs, the deadline of each transaction is crucial. In time-sensitive cases, the transaction should be settled before the deadline, but few studies focus on the deadline of the transaction. Most studies focus on routing in PCNs to improve throughput, but generally ignore the latency in multi-path routing. Although the study in~\cite{Papadis2022Single} considers the transaction deadlines in the single-hop transaction scheduling problem, it fixes the number of transactions. Authors in~\cite{Qiu2022A} investigate privacy-preserving transaction scheduling, considering that the key to payment settlement for PCNs is to find a path with sufficient balance, but the deadline for transactions is not taken into account. Authors in~\cite{sivaraman2020high} divide transactions into different transaction units and control the number of transaction units allocated to each path by designing congestion control algorithms. These transaction units can then be sent at different rates across different paths, enabling high throughput routing. However, a successful transaction requires settling all its transaction units, meaning the latency of a transaction depends on the settlement time of the last transaction unit. Furthermore, this study does not consider the transaction deadline.

\indent To achieve high throughput without fixing the number of transactions, how to ensure that the completion time of multi-path transactions is less than the maximum deadline can be a direction for future research, by comprehensively considering transaction sending rates, reliability, congestion levels and transaction demands of different nodes.

\subsection{Lightweight Off-chain Storage}

\indent To alleviate the storage limitations of blockchain, most researchers choose to use off-chain storage to extend the performance of blockchain. Off-chain storage stores data completely outside the blockchain on external storage media, such as cloud services, IPFS and local servers. However, off-chain storage faces challenges such as centralization risks, storage costs, data access efficiency, data privacy and security. Compared to traditional off-chain storage solutions, lightweight off-chain storage provides a potential solution that improves data reliability and security through data sharding and data redundancy, while synchronizing data to the blockchain only when necessary. 

\indent The massive amount of data can make distributed ledgers cumbersome, and lightweight storage is a solution for efficient storage. However, current research generally focus on storing lightweight index or transaction information on-chain, and using local servers to store massive amounts of data off-chain, ignoring lightweight off-chain storage~\cite{li2022efficient}. It is a critical challenge for off-chain networks. In the future, lightweight technologies need to be leveraged to minimize storage costs and bandwidth requirements for data transmission, while ensuring both performance and security requirements. In addition, while the amount of data stored and accessed off-chain increases, it is important to ensure storage sustainability while avoiding centralized control and excessive costs.

\subsection{Balance between Transaction Throughput and Privacy Risks}

\indent In off-chain networks, increasing the throughput of transactions means increasing the efficiency of off-chain networks, but may sacrifice some privacy because transaction data needs to be disseminated and processed extensively. Conversely, increased privacy can impact transaction throughput and efficiency of off-chain networks. In existing PCNs, transactions are processed independently as soon as they arrive. If the transaction amount exceeds the available balance on each path, this may result in transaction failure and affect transaction throughput. Due to the heavy-tailed nature of transactions, a large portion of transactions may fail~\cite{Malavolt2017Concurrency}. Even if nodes have sufficient balances, different transaction flows can interfere with each other and cause deadlocks. Additionally, different transmission rates in opposite directions and transaction delays can significantly impact on the capacity of PCNs. To address this issue, efficient congestion control technologies need to be designed and implemented.

\indent Balancing transaction throughput and privacy risks in PCNs is crucial. Many studies that aim to increase transaction throughput raise concerns about user privacy and security. Although the authors in~\cite{Luo2022Learning} improve transaction throughput through priority awareness and introduce regulatory factors to meet the trade-off between throughput and forwarding costs, they do not consider the trade-off between transaction throughput and privacy. In achieving high transaction throughput, it is necessary to take appropriate measures to protect user privacy. Existing solutions for protecting privacy in off-chain networks do not work well in maintaining high throughput transaction processing speeds, and implementing these techniques also consumes a lot of computing and storage resources.

\section{Conclusion}\label{six}

\indent In this article, we provide a comprehensive review of the research on off-chain networks, focusing on their framework, issues, key technologies and solutions. Firstly, we introduce the background of off-chain networks, such as motivation, architecture, platforms, and application scenarios of off-chain networks. We then study and analyze the issues deserving to be considered in off-chain networks. In addition, we discuss several technologies and corresponding solutions to the aforementioned issues. Finally, we point out some research challenges and future research directions. We believe that this article will promote the further development of off-chain networks.

\bibliographystyle{ACM-Reference-Format}
\bibliography{huhaob}


\begin{thebibliography}{91}


\ifx \showCODEN    \undefined \def \showCODEN     #1{\unskip}     \fi
\ifx \showDOI      \undefined \def \showDOI       #1{#1}\fi
\ifx \showISBNx    \undefined \def \showISBNx     #1{\unskip}     \fi
\ifx \showISBNxiii \undefined \def \showISBNxiii  #1{\unskip}     \fi
\ifx \showISSN     \undefined \def \showISSN      #1{\unskip}     \fi
\ifx \showLCCN     \undefined \def \showLCCN      #1{\unskip}     \fi
\ifx \shownote     \undefined \def \shownote      #1{#1}          \fi
\ifx \showarticletitle \undefined \def \showarticletitle #1{#1}   \fi
\ifx \showURL      \undefined \def \showURL       {\relax}        \fi
\providecommand\bibfield[2]{#2}
\providecommand\bibinfo[2]{#2}
\providecommand\natexlab[1]{#1}
\providecommand\showeprint[2][]{arXiv:#2}

\bibitem[Alladi et~al\mbox{.}(2022)]%
        {Alladi2022A}
\bibfield{author}{\bibinfo{person}{Tejasvi Alladi}, \bibinfo{person}{Vinay
  Chamola}, \bibinfo{person}{Nishad Sahu}, \bibinfo{person}{Vishnu Venkatesh},
  \bibinfo{person}{Adit Goyal}, {and} \bibinfo{person}{Mohsen Guizani}.}
  \bibinfo{year}{2022}\natexlab{}.
\newblock \showarticletitle{{A Comprehensive Survey on the Applications of
  Blockchain for Securing Vehicular Networks}}.
\newblock \bibinfo{journal}{\emph{IEEE Communications Surveys \& Tutorials}}
  \bibinfo{volume}{24}, \bibinfo{number}{2} (\bibinfo{year}{2022}),
  \bibinfo{pages}{1212--1239}.
\newblock
\urldef\tempurl%
\url{https://doi.org/10.1109/COMST.2022.3160925}
\showDOI{\tempurl}


\bibitem[Arbabi et~al\mbox{.}(2023)]%
        {Arbabi2023A}
\bibfield{author}{\bibinfo{person}{Mohammad~Salar Arbabi},
  \bibinfo{person}{Chhagan Lal}, \bibinfo{person}{Narasimha~Raghavan
  Veeraragavan}, \bibinfo{person}{Dusica Marijan}, \bibinfo{person}{Jan~F.
  Nygård}, {and} \bibinfo{person}{Roman Vitenberg}.}
  \bibinfo{year}{2023}\natexlab{}.
\newblock \showarticletitle{{A Survey on Blockchain for Healthcare: Challenges,
  Benefits, and Future Directions}}.
\newblock \bibinfo{journal}{\emph{IEEE Communications Surveys \& Tutorials}}
  \bibinfo{volume}{25}, \bibinfo{number}{1} (\bibinfo{year}{2023}),
  \bibinfo{pages}{386--424}.
\newblock
\urldef\tempurl%
\url{https://doi.org/10.1109/COMST.2022.3224644}
\showDOI{\tempurl}


\bibitem[Asheralieva and Niyato(2021)]%
        {asheralieva2019learning}
\bibfield{author}{\bibinfo{person}{Alia Asheralieva} {and}
  \bibinfo{person}{Dusit Niyato}.} \bibinfo{year}{2021}\natexlab{}.
\newblock \showarticletitle{{Learning-Based Mobile Edge Computing Resource
  Management to Support Public Blockchain Networks}}.
\newblock \bibinfo{journal}{\emph{IEEE Transactions on Mobile Computing}}
  \bibinfo{volume}{20}, \bibinfo{number}{3} (\bibinfo{year}{2021}),
  \bibinfo{pages}{1092--1109}.
\newblock
\urldef\tempurl%
\url{https://doi.org/10.1109/TMC.2019.2959772}
\showDOI{\tempurl}


\bibitem[Azbeg et~al\mbox{.}(2022)]%
        {Azbeg2022Access}
\bibfield{author}{\bibinfo{person}{Kebira Azbeg}, \bibinfo{person}{Ouail
  Ouchetto}, {and} \bibinfo{person}{Said~Jai Andaloussi}.}
  \bibinfo{year}{2022}\natexlab{}.
\newblock \showarticletitle{{Access Control and Privacy-Preserving
  Blockchain-Based System for Diseases Management}}.
\newblock \bibinfo{journal}{\emph{IEEE Transactions on Computational Social
  Systems, DOI:{10.1109/TCSS.2022.3186945}}} (\bibinfo{year}{2022}),
  \bibinfo{pages}{1--13}.
\newblock


\bibitem[Cai et~al\mbox{.}(2021)]%
        {Cai2019Towards}
\bibfield{author}{\bibinfo{person}{Chengjun Cai}, \bibinfo{person}{Yifeng
  Zheng}, \bibinfo{person}{Yuefeng Du}, \bibinfo{person}{Zhan Qin}, {and}
  \bibinfo{person}{Cong Wang}.} \bibinfo{year}{2021}\natexlab{}.
\newblock \showarticletitle{{Towards Private, Robust, and Verifiable
  Crowdsensing Systems via Public Blockchains}}.
\newblock \bibinfo{journal}{\emph{IEEE Transactions on Dependable and Secure
  Computing}} \bibinfo{volume}{18}, \bibinfo{number}{4} (\bibinfo{year}{2021}),
  \bibinfo{pages}{1893--1907}.
\newblock
\urldef\tempurl%
\url{https://doi.org/10.1109/TDSC.2019.2941481}
\showDOI{\tempurl}


\bibitem[Cao et~al\mbox{.}(2023)]%
        {Cao2023Blockchain}
\bibfield{author}{\bibinfo{person}{Bin Cao}, \bibinfo{person}{Zixin Wang},
  \bibinfo{person}{Long Zhang}, \bibinfo{person}{Daquan Feng},
  \bibinfo{person}{Mugen Peng}, \bibinfo{person}{Lei Zhang}, {and}
  \bibinfo{person}{Zhu Han}.} \bibinfo{year}{2023}\natexlab{}.
\newblock \showarticletitle{{Blockchain Systems, Technologies, and
  Applications: A Methodology Perspective}}.
\newblock \bibinfo{journal}{\emph{IEEE Communications Surveys \& Tutorials}}
  \bibinfo{volume}{25}, \bibinfo{number}{1} (\bibinfo{year}{2023}),
  \bibinfo{pages}{353--385}.
\newblock
\urldef\tempurl%
\url{https://doi.org/10.1109/COMST.2022.3204702}
\showDOI{\tempurl}


\bibitem[Chai et~al\mbox{.}(2021)]%
        {Chai2021Hierarchical}
\bibfield{author}{\bibinfo{person}{Haoye Chai}, \bibinfo{person}{Supeng Leng},
  \bibinfo{person}{Yijin Chen}, {and} \bibinfo{person}{Ke Zhang}.}
  \bibinfo{year}{2021}\natexlab{}.
\newblock \showarticletitle{{A Hierarchical Blockchain-Enabled Federated
  Learning Algorithm for Knowledge Sharing in Internet of Vehicles}}.
\newblock \bibinfo{journal}{\emph{IEEE Transactions on Intelligent
  Transportation Systems}} \bibinfo{volume}{22}, \bibinfo{number}{7}
  (\bibinfo{year}{2021}), \bibinfo{pages}{3975--3986}.
\newblock
\urldef\tempurl%
\url{https://doi.org/10.1109/TITS.2020.3002712}
\showDOI{\tempurl}


\bibitem[Chen et~al\mbox{.}(2019a)]%
        {Chen2019CVCG}
\bibfield{author}{\bibinfo{person}{Chen Chen}, \bibinfo{person}{Jinna Hu},
  \bibinfo{person}{Tie Qiu}, \bibinfo{person}{Mohammed Atiquzzaman}, {and}
  \bibinfo{person}{Zhiyuan Ren}.} \bibinfo{year}{2019}\natexlab{a}.
\newblock \showarticletitle{{CVCG: Cooperative V2V-Aided Transmission Scheme
  Based on Coalitional Game for Popular Content Distribution in Vehicular
  Ad-Hoc Networks}}.
\newblock \bibinfo{journal}{\emph{IEEE Transactions on Mobile Computing}}
  \bibinfo{volume}{18}, \bibinfo{number}{12} (\bibinfo{year}{2019}),
  \bibinfo{pages}{2811--2828}.
\newblock
\urldef\tempurl%
\url{https://doi.org/10.1109/TMC.2018.2883312}
\showDOI{\tempurl}


\bibitem[Chen et~al\mbox{.}(2019b)]%
        {Chen2019A}
\bibfield{author}{\bibinfo{person}{Chuan Chen}, \bibinfo{person}{Jiajing Wu},
  \bibinfo{person}{Hui Lin}, \bibinfo{person}{Wuhui Chen}, {and}
  \bibinfo{person}{Zibin Zheng}.} \bibinfo{year}{2019}\natexlab{b}.
\newblock \showarticletitle{{A Secure and Efficient Blockchain-Based Data
  Trading Approach for Internet of Vehicles}}.
\newblock \bibinfo{journal}{\emph{IEEE Transactions on Vehicular Technology}}
  \bibinfo{volume}{68}, \bibinfo{number}{9} (\bibinfo{year}{2019}),
  \bibinfo{pages}{9110--9121}.
\newblock
\urldef\tempurl%
\url{https://doi.org/10.1109/TVT.2019.2927533}
\showDOI{\tempurl}


\bibitem[Chen et~al\mbox{.}(2021)]%
        {chen2021distributed}
\bibfield{author}{\bibinfo{person}{Mingzhe Chen}, \bibinfo{person}{Deniz
  Gündüz}, \bibinfo{person}{Kaibin Huang}, \bibinfo{person}{Walid Saad},
  \bibinfo{person}{Mehdi Bennis}, \bibinfo{person}{Aneta~Vulgarakis Feljan},
  {and} \bibinfo{person}{H.~Vincent Poor}.} \bibinfo{year}{2021}\natexlab{}.
\newblock \showarticletitle{{Distributed Learning in Wireless Networks: Recent
  Progress and Future Challenges}}.
\newblock \bibinfo{journal}{\emph{IEEE Journal on Selected Areas in
  Communications}} \bibinfo{volume}{39}, \bibinfo{number}{12}
  (\bibinfo{year}{2021}), \bibinfo{pages}{3579--3605}.
\newblock
\urldef\tempurl%
\url{https://doi.org/10.1109/JSAC.2021.3118346}
\showDOI{\tempurl}


\bibitem[Chen et~al\mbox{.}(2020)]%
        {Chen2020Toward}
\bibfield{author}{\bibinfo{person}{Wuhui Chen}, \bibinfo{person}{Yufei Chen},
  \bibinfo{person}{Xu Chen}, {and} \bibinfo{person}{Zibin Zheng}.}
  \bibinfo{year}{2020}\natexlab{}.
\newblock \showarticletitle{{Toward Secure Data Sharing for the IoV: A
  Quality-Driven Incentive Mechanism With On-Chain and Off-Chain Guarantees}}.
\newblock \bibinfo{journal}{\emph{IEEE Internet of Things Journal}}
  \bibinfo{volume}{7}, \bibinfo{number}{3} (\bibinfo{year}{2020}),
  \bibinfo{pages}{1625--1640}.
\newblock
\urldef\tempurl%
\url{https://doi.org/10.1109/JIOT.2019.2946611}
\showDOI{\tempurl}


\bibitem[Chen et~al\mbox{.}(2022b)]%
        {Chen2022Timeliness}
\bibfield{author}{\bibinfo{person}{Xianhao Chen}, \bibinfo{person}{Lan Zhang},
  \bibinfo{person}{Yawei Pang}, \bibinfo{person}{Bin Lin}, {and}
  \bibinfo{person}{Yuguang Fang}.} \bibinfo{year}{2022}\natexlab{b}.
\newblock \showarticletitle{{Timeliness-Aware Incentive Mechanism for Vehicular
  Crowdsourcing in Smart Cities}}.
\newblock \bibinfo{journal}{\emph{IEEE Transactions on Mobile Computing}}
  \bibinfo{volume}{21}, \bibinfo{number}{9} (\bibinfo{year}{2022}),
  \bibinfo{pages}{3373--3387}.
\newblock
\urldef\tempurl%
\url{https://doi.org/10.1109/TMC.2021.3052963}
\showDOI{\tempurl}


\bibitem[Chen et~al\mbox{.}(2022a)]%
        {Chen2022Multi}
\bibfield{author}{\bibinfo{person}{Yanjiao Chen}, \bibinfo{person}{Xuxian Li},
  \bibinfo{person}{Jian Zhang}, {and} \bibinfo{person}{Hongliang Bi}.}
  \bibinfo{year}{2022}\natexlab{a}.
\newblock \showarticletitle{{Multi-Party Payment Channel Network Based on Smart
  Contract}}.
\newblock \bibinfo{journal}{\emph{IEEE Transactions on Network and Service
  Management}} \bibinfo{volume}{19}, \bibinfo{number}{4}
  (\bibinfo{year}{2022}), \bibinfo{pages}{4847--4857}.
\newblock
\urldef\tempurl%
\url{https://doi.org/10.1109/TNSM.2022.3162592}
\showDOI{\tempurl}


\bibitem[Dotan et~al\mbox{.}(2021)]%
        {3453161}
\bibfield{author}{\bibinfo{person}{Maya Dotan}, \bibinfo{person}{Yvonne-Anne
  Pignolet}, \bibinfo{person}{Stefan Schmid}, \bibinfo{person}{Saar Tochner},
  {and} \bibinfo{person}{Aviv Zohar}.} \bibinfo{year}{2021}\natexlab{}.
\newblock \showarticletitle{Survey on Blockchain Networking: Context,
  State-of-the-Art, Challenges}.
\newblock \bibinfo{journal}{\emph{ACM Comput. Surv.}} \bibinfo{volume}{54},
  \bibinfo{number}{5}, Article \bibinfo{articleno}{107} (\bibinfo{date}{may}
  \bibinfo{year}{2021}), \bibinfo{numpages}{34}~pages.
\newblock
\showISSN{0360-0300}
\urldef\tempurl%
\url{https://doi.org/10.1145/3453161}
\showDOI{\tempurl}


\bibitem[Du et~al\mbox{.}(2022)]%
        {Du2022Anti-Collusion}
\bibfield{author}{\bibinfo{person}{Miao Du}, \bibinfo{person}{Peng Yang},
  \bibinfo{person}{Wen Tian}, {and} \bibinfo{person}{Zhu Han}.}
  \bibinfo{year}{2022}\natexlab{}.
\newblock \showarticletitle{{Anti-Collusion Multiparty Smart Contracts for
  Distributed Watchtowers in Payment Channel Networks}}.
\newblock \bibinfo{journal}{\emph{IEEE Journal on Selected Areas in
  Communications}} \bibinfo{volume}{40}, \bibinfo{number}{12}
  (\bibinfo{year}{2022}), \bibinfo{pages}{3600--3614}.
\newblock
\urldef\tempurl%
\url{https://doi.org/10.1109/JSAC.2022.3213355}
\showDOI{\tempurl}


\bibitem[Egger et~al\mbox{.}(2019)]%
        {Egger2019Atomic}
\bibfield{author}{\bibinfo{person}{Christoph Egger}, \bibinfo{person}{Pedro
  Moreno-Sanchez}, {and} \bibinfo{person}{Matteo Maffei}.}
  \bibinfo{year}{2019}\natexlab{}.
\newblock \showarticletitle{{Atomic Multi-Channel Updates with Constant
  Collateral in Bitcoin-Compatible Payment-Channel Networks}}. In
  \bibinfo{booktitle}{\emph{ACM SIGSAC Conference on Computer and
  Communications Security (CCS)}}. \bibinfo{pages}{801--815}.
\newblock


\bibitem[Fahmideh et~al\mbox{.}(2022)]%
        {10.1145/3530813}
\bibfield{author}{\bibinfo{person}{Mahdi Fahmideh}, \bibinfo{person}{John
  Grundy}, \bibinfo{person}{Aakash Ahmad}, \bibinfo{person}{Jun Shen},
  \bibinfo{person}{Jun Yan}, \bibinfo{person}{Davoud Mougouei},
  \bibinfo{person}{Peng Wang}, \bibinfo{person}{Aditya Ghose},
  \bibinfo{person}{Anuradha Gunawardana}, \bibinfo{person}{Uwe Aickelin}, {and}
  \bibinfo{person}{Babak Abedin}.} \bibinfo{year}{2022}\natexlab{}.
\newblock \showarticletitle{{Engineering Blockchain-Based Software Systems:
  Foundations, Survey, and Future Directions}}.
\newblock \bibinfo{journal}{\emph{Comput. Surveys}} \bibinfo{volume}{55},
  \bibinfo{number}{6} (\bibinfo{year}{2022}), \bibinfo{pages}{1--44}.
\newblock
\urldef\tempurl%
\url{https://doi.org/10.1145/3530813}
\showDOI{\tempurl}


\bibitem[Fan et~al\mbox{.}(2021)]%
        {Fan2020Hybrid}
\bibfield{author}{\bibinfo{person}{Sizheng Fan}, \bibinfo{person}{Hongbo
  Zhang}, \bibinfo{person}{Yuchen Zeng}, {and} \bibinfo{person}{Wei Cai}.}
  \bibinfo{year}{2021}\natexlab{}.
\newblock \showarticletitle{{Hybrid Blockchain-Based Resource Trading System
  for Federated Learning in Edge Computing}}.
\newblock \bibinfo{journal}{\emph{IEEE Internet of Things Journal}}
  \bibinfo{volume}{8}, \bibinfo{number}{4} (\bibinfo{year}{2021}),
  \bibinfo{pages}{2252--2264}.
\newblock
\urldef\tempurl%
\url{https://doi.org/10.1109/JIOT.2020.3028101}
\showDOI{\tempurl}


\bibitem[Gabay et~al\mbox{.}(2020)]%
        {Gabay2020Privacy}
\bibfield{author}{\bibinfo{person}{David Gabay}, \bibinfo{person}{Kemal
  Akkaya}, {and} \bibinfo{person}{Mumin Cebe}.}
  \bibinfo{year}{2020}\natexlab{}.
\newblock \showarticletitle{{Privacy-Preserving Authentication Scheme for
  Connected Electric Vehicles Using Blockchain and Zero Knowledge Proofs}}.
\newblock \bibinfo{journal}{\emph{IEEE Transactions on Vehicular Technology}}
  \bibinfo{volume}{69}, \bibinfo{number}{6} (\bibinfo{year}{2020}),
  \bibinfo{pages}{5760--5772}.
\newblock
\urldef\tempurl%
\url{https://doi.org/10.1109/TVT.2020.2977361}
\showDOI{\tempurl}


\bibitem[Gao et~al\mbox{.}(2018)]%
        {Gao2018Truthful}
\bibfield{author}{\bibinfo{person}{Guoju Gao}, \bibinfo{person}{Mingjun Xiao},
  \bibinfo{person}{Jie Wu}, \bibinfo{person}{Liusheng Huang}, {and}
  \bibinfo{person}{Chang Hu}.} \bibinfo{year}{2018}\natexlab{}.
\newblock \showarticletitle{{Truthful Incentive Mechanism for Nondeterministic
  Crowdsensing with Vehicles}}.
\newblock \bibinfo{journal}{\emph{IEEE Transactions on Mobile Computing}}
  \bibinfo{volume}{17}, \bibinfo{number}{12} (\bibinfo{year}{2018}),
  \bibinfo{pages}{2982--2997}.
\newblock
\urldef\tempurl%
\url{https://doi.org/10.1109/TMC.2018.2829506}
\showDOI{\tempurl}


\bibitem[Ge et~al\mbox{.}(2023)]%
        {Ge2023Magma}
\bibfield{author}{\bibinfo{person}{Zhonghui Ge}, \bibinfo{person}{Yi Zhang},
  \bibinfo{person}{Yu Long}, {and} \bibinfo{person}{Dawu Gu}.}
  \bibinfo{year}{2023}\natexlab{}.
\newblock \showarticletitle{{Magma: Robust and Flexible Multi-Party Payment
  Channel}}.
\newblock \bibinfo{journal}{\emph{IEEE Transactions on Dependable and Secure
  Computing, DOI:{10.1109/TDSC.2023.3238332}}} (\bibinfo{year}{2023}),
  \bibinfo{pages}{1--18}.
\newblock


\bibitem[Ghafoorian et~al\mbox{.}(2019)]%
        {ghafoorian2018thorough}
\bibfield{author}{\bibinfo{person}{Mahdi Ghafoorian}, \bibinfo{person}{Dariush
  Abbasinezhad-Mood}, {and} \bibinfo{person}{Hassan Shakeri}.}
  \bibinfo{year}{2019}\natexlab{}.
\newblock \showarticletitle{{A Thorough Trust and Reputation Based RBAC Model
  for Secure Data Storage in the Cloud}}.
\newblock \bibinfo{journal}{\emph{IEEE Transactions on Parallel and Distributed
  Systems}} \bibinfo{volume}{30}, \bibinfo{number}{4} (\bibinfo{year}{2019}),
  \bibinfo{pages}{778--788}.
\newblock
\urldef\tempurl%
\url{https://doi.org/10.1109/TPDS.2018.2870652}
\showDOI{\tempurl}


\bibitem[Green and Miers(2017)]%
        {Green2017Bolt}
\bibfield{author}{\bibinfo{person}{Matthew Green} {and} \bibinfo{person}{Ian
  Miers}.} \bibinfo{year}{2017}\natexlab{}.
\newblock \showarticletitle{{Bolt: Anonymous Payment Channels for Decentralized
  Currencies}}. In \bibinfo{booktitle}{\emph{ACM SIGSAC Conference on Computer
  and Communications Security (CCS)}}. \bibinfo{pages}{473--489}.
\newblock


\bibitem[Hao et~al\mbox{.}(2022)]%
        {Hao2022Smart}
\bibfield{author}{\bibinfo{person}{Jialu Hao}, \bibinfo{person}{Cheng Huang},
  \bibinfo{person}{Wenjuan Tang}, \bibinfo{person}{Yang Zhang}, {and}
  \bibinfo{person}{Shuai Yuan}.} \bibinfo{year}{2022}\natexlab{}.
\newblock \showarticletitle{{Smart Contract-Based Access Control Through
  Off-Chain Signature and On-Chain Evaluation}}.
\newblock \bibinfo{journal}{\emph{IEEE Transactions on Circuits and Systems II:
  Express Briefs}} \bibinfo{volume}{69}, \bibinfo{number}{4}
  (\bibinfo{year}{2022}), \bibinfo{pages}{2221--2225}.
\newblock
\urldef\tempurl%
\url{https://doi.org/10.1109/TCSII.2021.3125500}
\showDOI{\tempurl}


\bibitem[He et~al\mbox{.}(2022)]%
        {He2022Bift}
\bibfield{author}{\bibinfo{person}{Ying He}, \bibinfo{person}{Ke Huang},
  \bibinfo{person}{Guangzheng Zhang}, \bibinfo{person}{F.~Richard Yu},
  \bibinfo{person}{Jianyong Chen}, {and} \bibinfo{person}{Jianqiang Li}.}
  \bibinfo{year}{2022}\natexlab{}.
\newblock \showarticletitle{{Bift: A Blockchain-Based Federated Learning System
  for Connected and Autonomous Vehicles}}.
\newblock \bibinfo{journal}{\emph{IEEE Internet of Things Journal}}
  \bibinfo{volume}{9}, \bibinfo{number}{14} (\bibinfo{year}{2022}),
  \bibinfo{pages}{12311--12322}.
\newblock
\urldef\tempurl%
\url{https://doi.org/10.1109/JIOT.2021.3135342}
\showDOI{\tempurl}


\bibitem[Heilman et~al\mbox{.}(2017)]%
        {Heilman2017Tumblebit}
\bibfield{author}{\bibinfo{person}{Ethan Heilman}, \bibinfo{person}{Leen
  Alshenibr}, \bibinfo{person}{Foteini Baldimtsi}, \bibinfo{person}{Alessandra
  Scafuro}, {and} \bibinfo{person}{Sharon Goldberg}.}
  \bibinfo{year}{2017}\natexlab{}.
\newblock \showarticletitle{{TumbleBit: An Untrusted Bitcoin-Compatible
  Anonymous Payment Hub}}. In \bibinfo{booktitle}{\emph{Network and Distributed
  System Security Symposium (NDSS), DOI:{10.14722/NDSS.2017.23086}}}.
\newblock


\bibitem[Jayabalan and Jeyanthi(2022)]%
        {jayabalan2022scalable}
\bibfield{author}{\bibinfo{person}{Jayapriya Jayabalan} {and}
  \bibinfo{person}{N Jeyanthi}.} \bibinfo{year}{2022}\natexlab{}.
\newblock \showarticletitle{{Scalable Blockchain Model using Off-chain IPFS
  Storage for Healthcare Data Security and Privacy}}.
\newblock \bibinfo{journal}{\emph{J. Parallel and Distrib. Comput.}}
  \bibinfo{volume}{164} (\bibinfo{year}{2022}), \bibinfo{pages}{152--167}.
\newblock


\bibitem[Kang et~al\mbox{.}(2019)]%
        {Kang2019Blockchain}
\bibfield{author}{\bibinfo{person}{Jiawen Kang}, \bibinfo{person}{Rong Yu},
  \bibinfo{person}{Xumin Huang}, \bibinfo{person}{Maoqiang Wu},
  \bibinfo{person}{Sabita Maharjan}, \bibinfo{person}{Shengli Xie}, {and}
  \bibinfo{person}{Yan Zhang}.} \bibinfo{year}{2019}\natexlab{}.
\newblock \showarticletitle{{Blockchain for Secure and Efficient Data Sharing
  in Vehicular Edge Computing and Networks}}.
\newblock \bibinfo{journal}{\emph{IEEE Internet of Things Journal}}
  \bibinfo{volume}{6}, \bibinfo{number}{3} (\bibinfo{year}{2019}),
  \bibinfo{pages}{4660--4670}.
\newblock
\urldef\tempurl%
\url{https://doi.org/10.1109/JIOT.2018.2875542}
\showDOI{\tempurl}


\bibitem[Khalil and Gervais(2017)]%
        {khalil2017revive}
\bibfield{author}{\bibinfo{person}{Rami Khalil} {and} \bibinfo{person}{Arthur
  Gervais}.} \bibinfo{year}{2017}\natexlab{}.
\newblock \showarticletitle{{Revive: Rebalancing Off-Blockchain Payment
  Networks}}. In \bibinfo{booktitle}{\emph{ACM SIGSAC Conference on Computer
  and Communications Security (CCS)}}. \bibinfo{pages}{439--453}.
\newblock


\bibitem[Khan et~al\mbox{.}(2021)]%
        {9460016}
\bibfield{author}{\bibinfo{person}{Latif~U. Khan}, \bibinfo{person}{Walid
  Saad}, \bibinfo{person}{Zhu Han}, \bibinfo{person}{Ekram Hossain}, {and}
  \bibinfo{person}{Choong~Seon Hong}.} \bibinfo{year}{2021}\natexlab{}.
\newblock \showarticletitle{{Federated Learning for Internet of Things: Recent
  Advances, Taxonomy, and Open Challenges}}.
\newblock \bibinfo{journal}{\emph{IEEE Communications Surveys \& Tutorials}}
  \bibinfo{volume}{23}, \bibinfo{number}{3} (\bibinfo{year}{2021}),
  \bibinfo{pages}{1759--1799}.
\newblock
\urldef\tempurl%
\url{https://doi.org/10.1109/COMST.2021.3090430}
\showDOI{\tempurl}


\bibitem[Kim et~al\mbox{.}(2018)]%
        {kim2018sgx}
\bibfield{author}{\bibinfo{person}{Seongmin Kim}, \bibinfo{person}{Juhyeng
  Han}, \bibinfo{person}{Jaehyeong Ha}, \bibinfo{person}{Taesoo Kim}, {and}
  \bibinfo{person}{Dongsu Han}.} \bibinfo{year}{2018}\natexlab{}.
\newblock \showarticletitle{{SGX-Tor: A Secure and Practical Tor Anonymity
  Network With SGX Enclaves}}.
\newblock \bibinfo{journal}{\emph{IEEE/ACM Transactions on Networking}}
  \bibinfo{volume}{26}, \bibinfo{number}{5} (\bibinfo{year}{2018}),
  \bibinfo{pages}{2174--2187}.
\newblock
\urldef\tempurl%
\url{https://doi.org/10.1109/TNET.2018.2868054}
\showDOI{\tempurl}


\bibitem[Kurt et~al\mbox{.}(2023)]%
        {Kurt2023A}
\bibfield{author}{\bibinfo{person}{Ahmet Kurt}, \bibinfo{person}{Enes Erdin},
  \bibinfo{person}{Kemal Akkaya}, \bibinfo{person}{Selcuk Uluagac}, {and}
  \bibinfo{person}{Mumin Cebe}.} \bibinfo{year}{2023}\natexlab{}.
\newblock \showarticletitle{{D-LNBot: A Scalable, Cost-Free and Covert Hybrid
  Botnet on Bitcoin's Lightning Network}}.
\newblock \bibinfo{journal}{\emph{IEEE Transactions on Dependable and Secure
  Computing, DOI:{10.1109/TDSC.2023.3300738}}} (\bibinfo{year}{2023}),
  \bibinfo{pages}{1--18}.
\newblock


\bibitem[Kus~Khalilov and Levi(2018)]%
        {Kus2018ASurvey}
\bibfield{author}{\bibinfo{person}{Merve~Can Kus~Khalilov} {and}
  \bibinfo{person}{Albert Levi}.} \bibinfo{year}{2018}\natexlab{}.
\newblock \showarticletitle{{A Survey on Anonymity and Privacy in Bitcoin-Like
  Digital Cash Systems}}.
\newblock \bibinfo{journal}{\emph{IEEE Communications Surveys \& Tutorials}}
  \bibinfo{volume}{20}, \bibinfo{number}{3} (\bibinfo{year}{2018}),
  \bibinfo{pages}{2543--2585}.
\newblock
\urldef\tempurl%
\url{https://doi.org/10.1109/COMST.2018.2818623}
\showDOI{\tempurl}


\bibitem[Li et~al\mbox{.}(2022a)]%
        {li2022efficient}
\bibfield{author}{\bibinfo{person}{Chaoyang Li}, \bibinfo{person}{Mianxiong
  Dong}, \bibinfo{person}{Jian Li}, \bibinfo{person}{Gang Xu},
  \bibinfo{person}{Xiu-Bo Chen}, \bibinfo{person}{Wen Liu}, {and}
  \bibinfo{person}{Kaoru Ota}.} \bibinfo{year}{2022}\natexlab{a}.
\newblock \showarticletitle{{Efficient Medical Big Data Management With
  Keyword-Searchable Encryption in Healthchain}}.
\newblock \bibinfo{journal}{\emph{IEEE Systems Journal}} \bibinfo{volume}{16},
  \bibinfo{number}{4} (\bibinfo{year}{2022}), \bibinfo{pages}{5521--5532}.
\newblock
\urldef\tempurl%
\url{https://doi.org/10.1109/JSYST.2022.3173538}
\showDOI{\tempurl}


\bibitem[Li et~al\mbox{.}(2022b)]%
        {Li2022Compact}
\bibfield{author}{\bibinfo{person}{Zhenni Li}, \bibinfo{person}{Wensheng Su},
  \bibinfo{person}{Minrui Xu}, \bibinfo{person}{Rong Yu},
  \bibinfo{person}{Dusit Niyato}, {and} \bibinfo{person}{Shengli Xie}.}
  \bibinfo{year}{2022}\natexlab{b}.
\newblock \showarticletitle{{Compact Learning Model for Dynamic Off-chain
  Routing in Blockchain-Based IoT}}.
\newblock \bibinfo{journal}{\emph{IEEE Journal on Selected Areas in
  Communications}} \bibinfo{volume}{40}, \bibinfo{number}{12}
  (\bibinfo{year}{2022}), \bibinfo{pages}{3615--3630}.
\newblock
\urldef\tempurl%
\url{https://doi.org/10.1109/JSAC.2022.3213283}
\showDOI{\tempurl}


\bibitem[Liang et~al\mbox{.}(2022)]%
        {Liang2022PDPChain}
\bibfield{author}{\bibinfo{person}{Wei Liang}, \bibinfo{person}{Yang Yang},
  \bibinfo{person}{Ce Yang}, \bibinfo{person}{Yonghua Hu},
  \bibinfo{person}{Songyou Xie}, \bibinfo{person}{Kuan-Ching Li}, {and}
  \bibinfo{person}{Jiannong Cao}.} \bibinfo{year}{2022}\natexlab{}.
\newblock \showarticletitle{{PDPChain: A Consortium Blockchain-Based Privacy
  Protection Scheme for Personal Data}}.
\newblock \bibinfo{journal}{\emph{IEEE Transactions on Reliability,
  DOI:{10.1109/TR.2022.3190932}}} (\bibinfo{year}{2022}),
  \bibinfo{pages}{1--13}.
\newblock


\bibitem[Lin et~al\mbox{.}(2021)]%
        {lin2021model}
\bibfield{author}{\bibinfo{person}{Yijing Lin}, \bibinfo{person}{Zhipeng Gao},
  \bibinfo{person}{Kaile Xiao}, \bibinfo{person}{Qian Wang},
  \bibinfo{person}{Zijia Mo}, \bibinfo{person}{Yang Yang},
  \bibinfo{person}{Lanlan Rui}, \bibinfo{person}{Haisheng Guo}, {and}
  \bibinfo{person}{Dezheng Wang}.} \bibinfo{year}{2021}\natexlab{}.
\newblock \showarticletitle{{A Model Training Mechanism based on Onchain and
  Offchain Collaboration for Edge Computing}}. In
  \bibinfo{booktitle}{\emph{IEEE International Conference on Communications
  (ICC)}}. \bibinfo{pages}{1--6}.
\newblock


\bibitem[Liu et~al\mbox{.}(2022a)]%
        {Liu2022Extending}
\bibfield{author}{\bibinfo{person}{Chunchi Liu}, \bibinfo{person}{Hechuan Guo},
  \bibinfo{person}{Minghui Xu}, \bibinfo{person}{Shengling Wang},
  \bibinfo{person}{Dongxiao Yu}, \bibinfo{person}{Jiguo Yu}, {and}
  \bibinfo{person}{Xiuzhen Cheng}.} \bibinfo{year}{2022}\natexlab{a}.
\newblock \showarticletitle{{Extending On-Chain Trust to Off-Chain –
  Trustworthy Blockchain Data Collection using Trusted Execution Environment
  (TEE)}}.
\newblock \bibinfo{journal}{\emph{IEEE Trans. Comput.}} \bibinfo{volume}{71},
  \bibinfo{number}{12} (\bibinfo{year}{2022}), \bibinfo{pages}{3268--3280}.
\newblock
\urldef\tempurl%
\url{https://doi.org/10.1109/TC.2022.3148379}
\showDOI{\tempurl}


\bibitem[Liu et~al\mbox{.}(2022b)]%
        {Liu2022Blockchain}
\bibfield{author}{\bibinfo{person}{Dongxiao Liu}, \bibinfo{person}{Huaqing Wu},
  \bibinfo{person}{Cheng Huang}, \bibinfo{person}{Jianbing Ni}, {and}
  \bibinfo{person}{Xuemin Shen}.} \bibinfo{year}{2022}\natexlab{b}.
\newblock \showarticletitle{{Blockchain-Based Credential Management for
  Anonymous Authentication in SAGVN}}.
\newblock \bibinfo{journal}{\emph{IEEE Journal on Selected Areas in
  Communications}} \bibinfo{volume}{40}, \bibinfo{number}{10}
  (\bibinfo{year}{2022}), \bibinfo{pages}{3104--3116}.
\newblock
\urldef\tempurl%
\url{https://doi.org/10.1109/JSAC.2022.3196091}
\showDOI{\tempurl}


\bibitem[Liu et~al\mbox{.}(2021)]%
        {liu2021blockchain}
\bibfield{author}{\bibinfo{person}{Yaowei Liu}, \bibinfo{person}{Zhiwei Wu},
  \bibinfo{person}{Yuchun Liu}, \bibinfo{person}{Wenbo Fan},
  \bibinfo{person}{Kai Wang}, {and} \bibinfo{person}{Lei Lin}.}
  \bibinfo{year}{2021}\natexlab{}.
\newblock \showarticletitle{{Blockchain Data Anti-Counterfeiting Sharing Model
  of Power Material Alliance Based on Attribute-Based Encryption}}. In
  \bibinfo{booktitle}{\emph{IEEE International Conference on Automation,
  Electronics and Electrical Engineering (AUTEEE)}}. \bibinfo{pages}{242--248}.
\newblock
\urldef\tempurl%
\url{https://doi.org/10.1109/AUTEEE52864.2021.9668791}
\showDOI{\tempurl}


\bibitem[Lu et~al\mbox{.}(2022)]%
        {Lu2022CoinLayering}
\bibfield{author}{\bibinfo{person}{Ning Lu}, \bibinfo{person}{Yuan Chang},
  \bibinfo{person}{Wenbo Shi}, {and} \bibinfo{person}{Kim-Kwang~Raymond Choo}.}
  \bibinfo{year}{2022}\natexlab{}.
\newblock \showarticletitle{{CoinLayering: An Efficient Coin Mixing Scheme for
  Large Scale Bitcoin Transactions}}.
\newblock \bibinfo{journal}{\emph{IEEE Transactions on Dependable and Secure
  Computing}} \bibinfo{volume}{19}, \bibinfo{number}{3} (\bibinfo{year}{2022}),
  \bibinfo{pages}{1974--1987}.
\newblock
\urldef\tempurl%
\url{https://doi.org/10.1109/TDSC.2020.3043366}
\showDOI{\tempurl}


\bibitem[Luo and Li(2022)]%
        {Luo2022Learning}
\bibfield{author}{\bibinfo{person}{Xiaofei Luo} {and} \bibinfo{person}{Peng
  Li}.} \bibinfo{year}{2022}\natexlab{}.
\newblock \showarticletitle{{Learning-Based Off-Chain Transaction Scheduling in
  Prioritized Payment Channel Networks}}.
\newblock \bibinfo{journal}{\emph{IEEE Journal on Selected Areas in
  Communications}} \bibinfo{volume}{40}, \bibinfo{number}{12}
  (\bibinfo{year}{2022}), \bibinfo{pages}{3589--3599}.
\newblock
\urldef\tempurl%
\url{https://doi.org/10.1109/JSAC.2022.3213333}
\showDOI{\tempurl}


\bibitem[Ma et~al\mbox{.}(2021)]%
        {Ma2021Research}
\bibfield{author}{\bibinfo{person}{Chao Ma}, \bibinfo{person}{Chen Song}, {and}
  \bibinfo{person}{Ma Yucong}.} \bibinfo{year}{2021}\natexlab{}.
\newblock \showarticletitle{{Research on Collaborative Security Based on
  On-blockchain and Off-blockchain in the Scenario of Power Data Acquisition}}.
  In \bibinfo{booktitle}{\emph{IEEE International Conference on Communication
  Technology (ICCT)}}. \bibinfo{pages}{876--881}.
\newblock
\urldef\tempurl%
\url{https://doi.org/10.1109/ICCT52962.2021.9658051}
\showDOI{\tempurl}


\bibitem[Malavolta et~al\mbox{.}(2016)]%
        {malavolta2016silentwhispers}
\bibfield{author}{\bibinfo{person}{Giulio Malavolta}, \bibinfo{person}{Pedro
  Moreno-Sanchez}, \bibinfo{person}{Aniket Kate}, {and} \bibinfo{person}{Matteo
  Maffei}.} \bibinfo{year}{2016}\natexlab{}.
\newblock \showarticletitle{{Silentwhispers: Enforcing Security and Privacy in
  Decentralized Credit Networks}}.
\newblock \bibinfo{journal}{\emph{Cryptology ePrint Archive}}
  (\bibinfo{year}{2016}).
\newblock


\bibitem[Malavolta et~al\mbox{.}(2017)]%
        {Malavolt2017Concurrency}
\bibfield{author}{\bibinfo{person}{Giulio Malavolta}, \bibinfo{person}{Pedro
  Moreno-Sanchez}, \bibinfo{person}{Aniket Kate}, \bibinfo{person}{Matteo
  Maffei}, {and} \bibinfo{person}{Srivatsan Ravi}.}
  \bibinfo{year}{2017}\natexlab{}.
\newblock \showarticletitle{{Concurrency and Privacy with Payment-Channel
  Networks}}. In \bibinfo{booktitle}{\emph{ACM SIGSAC Conference on Computer
  and Communications Security (CCS)}}. \bibinfo{pages}{455--471}.
\newblock


\bibitem[Mazumdar and Ruj(2023)]%
        {Mazumdar2022Cryptomaze}
\bibfield{author}{\bibinfo{person}{Subhra Mazumdar} {and}
  \bibinfo{person}{Sushmita Ruj}.} \bibinfo{year}{2023}\natexlab{}.
\newblock \showarticletitle{{CryptoMaze: Privacy-Preserving Splitting of
  Off-Chain Payments}}.
\newblock \bibinfo{journal}{\emph{IEEE Transactions on Dependable and Secure
  Computing}} \bibinfo{volume}{20}, \bibinfo{number}{2} (\bibinfo{year}{2023}),
  \bibinfo{pages}{1060--1073}.
\newblock
\urldef\tempurl%
\url{https://doi.org/10.1109/TDSC.2022.3148476}
\showDOI{\tempurl}


\bibitem[Nguyen et~al\mbox{.}(2021)]%
        {9415623}
\bibfield{author}{\bibinfo{person}{Dinh~C. Nguyen}, \bibinfo{person}{Ming
  Ding}, \bibinfo{person}{Pubudu~N. Pathirana}, \bibinfo{person}{Aruna
  Seneviratne}, \bibinfo{person}{Jun Li}, {and} \bibinfo{person}{H.
  Vincent~Poor}.} \bibinfo{year}{2021}\natexlab{}.
\newblock \showarticletitle{{Federated Learning for Internet of Things: A
  Comprehensive Survey}}.
\newblock \bibinfo{journal}{\emph{IEEE Communications Surveys \& Tutorials}}
  \bibinfo{volume}{23}, \bibinfo{number}{3} (\bibinfo{year}{2021}),
  \bibinfo{pages}{1622--1658}.
\newblock
\urldef\tempurl%
\url{https://doi.org/10.1109/COMST.2021.3075439}
\showDOI{\tempurl}


\bibitem[Ning et~al\mbox{.}(2022a)]%
        {Ning2022SmartGrids}
\bibfield{author}{\bibinfo{person}{Zhaolong Ning}, \bibinfo{person}{Handi
  Chen}, \bibinfo{person}{Xiaojie Wang}, \bibinfo{person}{Shupeng Wang}, {and}
  \bibinfo{person}{Lei Guo}.} \bibinfo{year}{2022}\natexlab{a}.
\newblock \showarticletitle{{Blockchain-Enabled Electrical Fault Inspection and
  Secure Transmission in 5G Smart Grids}}.
\newblock \bibinfo{journal}{\emph{IEEE Journal of Selected Topics in Signal
  Processing}} \bibinfo{volume}{16}, \bibinfo{number}{1}
  (\bibinfo{year}{2022}), \bibinfo{pages}{82--96}.
\newblock
\urldef\tempurl%
\url{https://doi.org/10.1109/JSTSP.2021.3120872}
\showDOI{\tempurl}


\bibitem[Ning et~al\mbox{.}(2022b)]%
        {Ning2022Blockchain}
\bibfield{author}{\bibinfo{person}{Zhaolong Ning}, \bibinfo{person}{Shouming
  Sun}, \bibinfo{person}{Xiaojie Wang}, \bibinfo{person}{Lei Guo},
  \bibinfo{person}{Song Guo}, \bibinfo{person}{Xiping Hu}, \bibinfo{person}{Bin
  Hu}, {and} \bibinfo{person}{Ricky Y.~K. Kwok}.}
  \bibinfo{year}{2022}\natexlab{b}.
\newblock \showarticletitle{{Blockchain-Enabled Intelligent Transportation
  Systems: A Distributed Crowdsensing Framework}}.
\newblock \bibinfo{journal}{\emph{IEEE Transactions on Mobile Computing}}
  \bibinfo{volume}{21}, \bibinfo{number}{12} (\bibinfo{year}{2022}),
  \bibinfo{pages}{4201--4217}.
\newblock
\urldef\tempurl%
\url{https://doi.org/10.1109/TMC.2021.3079984}
\showDOI{\tempurl}


\bibitem[Papadis and Tassiulas(2022)]%
        {Papadis2022Single}
\bibfield{author}{\bibinfo{person}{Nikolaos Papadis} {and}
  \bibinfo{person}{Leandros Tassiulas}.} \bibinfo{year}{2022}\natexlab{}.
\newblock \showarticletitle{{Payment Channel Networks: Single-Hop Scheduling
  for Throughput Maximization}}. In \bibinfo{booktitle}{\emph{IEEE Conference
  on Computer Communications (INFOCOM)}}. \bibinfo{pages}{900--909}.
\newblock
\urldef\tempurl%
\url{https://doi.org/10.1109/INFOCOM48880.2022.9796862}
\showDOI{\tempurl}


\bibitem[Pasdar et~al\mbox{.}(2023)]%
        {pasdar2023connect}
\bibfield{author}{\bibinfo{person}{Amirmohammad Pasdar},
  \bibinfo{person}{Young~Choon Lee}, {and} \bibinfo{person}{Zhongli Dong}.}
  \bibinfo{year}{2023}\natexlab{}.
\newblock \showarticletitle{{Connect API with Blockchain: A Survey on
  Blockchain Oracle Implementation}}.
\newblock \bibinfo{journal}{\emph{Comput. Surveys}} \bibinfo{volume}{55},
  \bibinfo{number}{10} (\bibinfo{year}{2023}), \bibinfo{pages}{1--39}.
\newblock


\bibitem[Pietrzak et~al\mbox{.}(2021)]%
        {pietrzak2021lightpir}
\bibfield{author}{\bibinfo{person}{Krzysztof Pietrzak}, \bibinfo{person}{Iosif
  Salem}, \bibinfo{person}{Stefan Schmid}, {and} \bibinfo{person}{Michelle
  Yeo}.} \bibinfo{year}{2021}\natexlab{}.
\newblock \showarticletitle{{LightPIR: Privacy-Preserving Route Discovery for
  Payment Channel Networks}}. In \bibinfo{booktitle}{\emph{IFIP Networking
  Conference (IFIP Networking)}}. \bibinfo{pages}{1--9}.
\newblock
\urldef\tempurl%
\url{https://doi.org/10.23919/IFIPNetworking52078.2021.9472205}
\showDOI{\tempurl}


\bibitem[Prihodko et~al\mbox{.}(2016)]%
        {prihodko2016flare}
\bibfield{author}{\bibinfo{person}{Pavel Prihodko}, \bibinfo{person}{Slava
  Zhigulin}, \bibinfo{person}{Mykola Sahno}, \bibinfo{person}{Aleksei
  Ostrovskiy}, {and} \bibinfo{person}{Olaoluwa Osuntokun}.}
  \bibinfo{year}{2016}\natexlab{}.
\newblock \showarticletitle{{Flare: An Approach to Routing in Lightning
  Network}}.
\newblock \bibinfo{journal}{\emph{White Paper}}  \bibinfo{volume}{144}
  (\bibinfo{year}{2016}).
\newblock


\bibitem[Qiu et~al\mbox{.}(2022)]%
        {Qiu2022A}
\bibfield{author}{\bibinfo{person}{Xiaoyu Qiu}, \bibinfo{person}{Wuhui Chen},
  \bibinfo{person}{Bingxin Tang}, \bibinfo{person}{Junyuan Liang},
  \bibinfo{person}{Hong-Ning Dai}, {and} \bibinfo{person}{Zibin Zheng}.}
  \bibinfo{year}{2022}\natexlab{}.
\newblock \showarticletitle{{A Distributed and Privacy-Aware High-Throughput
  Transaction Scheduling Approach for Scaling Blockchain}}.
\newblock \bibinfo{journal}{\emph{IEEE Transactions on Dependable and Secure
  Computing, DOI:{10.1109/TDSC.2022.3216571}}} (\bibinfo{year}{2022}),
  \bibinfo{pages}{1--15}.
\newblock


\bibitem[Roos et~al\mbox{.}(2017)]%
        {Roos2018Settling}
\bibfield{author}{\bibinfo{person}{S. Roos}, \bibinfo{person}{P.
  Moreno-Sanchez}, \bibinfo{person}{A. Kate}, {and} \bibinfo{person}{I.
  Goldberg}.} \bibinfo{year}{2017}\natexlab{}.
\newblock \showarticletitle{{Settling Payments Fast and Private: Efficient
  Decentralized Routing for Path-Based Transactions}}.
\newblock \bibinfo{journal}{\emph{arXiv preprint arXiv:1709.05748}}
  (\bibinfo{year}{2017}).
\newblock


\bibitem[Sebastian et~al\mbox{.}(2019)]%
        {Sebastian2019DER}
\bibfield{author}{\bibinfo{person}{D.~Jonathan Sebastian},
  \bibinfo{person}{Utkarsh Agrawal}, \bibinfo{person}{Ali Tamimi}, {and}
  \bibinfo{person}{Adam Hahn}.} \bibinfo{year}{2019}\natexlab{}.
\newblock \showarticletitle{{DER-TEE: Secure Distributed Energy Resource
  Operations Through Trusted Execution Environments}}.
\newblock \bibinfo{journal}{\emph{IEEE Internet of Things Journal}}
  \bibinfo{volume}{6}, \bibinfo{number}{4} (\bibinfo{year}{2019}),
  \bibinfo{pages}{6476--6486}.
\newblock
\urldef\tempurl%
\url{https://doi.org/10.1109/JIOT.2019.2909768}
\showDOI{\tempurl}


\bibitem[Shen et~al\mbox{.}(2020)]%
        {Shen2020Blockchain}
\bibfield{author}{\bibinfo{person}{Meng Shen}, \bibinfo{person}{Huisen Liu},
  \bibinfo{person}{Liehuang Zhu}, \bibinfo{person}{Ke Xu},
  \bibinfo{person}{Hongbo Yu}, \bibinfo{person}{Xiaojiang Du}, {and}
  \bibinfo{person}{Mohsen Guizani}.} \bibinfo{year}{2020}\natexlab{}.
\newblock \showarticletitle{{Blockchain-Assisted Secure Device Authentication
  for Cross-Domain Industrial IoT}}.
\newblock \bibinfo{journal}{\emph{IEEE Journal on Selected Areas in
  Communications}} \bibinfo{volume}{38}, \bibinfo{number}{5}
  (\bibinfo{year}{2020}), \bibinfo{pages}{942--954}.
\newblock
\urldef\tempurl%
\url{https://doi.org/10.1109/JSAC.2020.2980916}
\showDOI{\tempurl}


\bibitem[Sivaraman et~al\mbox{.}(2020)]%
        {sivaraman2020high}
\bibfield{author}{\bibinfo{person}{Vibhaalakshmi Sivaraman},
  \bibinfo{person}{Shaileshh~Bojja Venkatakrishnan}, \bibinfo{person}{Kathleen
  Ruan}, \bibinfo{person}{Parimarjan Negi}, \bibinfo{person}{Lei Yang},
  \bibinfo{person}{Radhika Mittal}, \bibinfo{person}{Giulia Fanti}, {and}
  \bibinfo{person}{Mohammad Alizadeh}.} \bibinfo{year}{2020}\natexlab{}.
\newblock \showarticletitle{{High Throughput Cryptocurrency Routing in Payment
  Channel Networks}}. In \bibinfo{booktitle}{\emph{USENIX Symposium on
  Networked Systems Design and Implementation (NSDI)}}.
  \bibinfo{pages}{777--796}.
\newblock


\bibitem[Su et~al\mbox{.}(2022)]%
        {Su2022LVBS}
\bibfield{author}{\bibinfo{person}{Zhou Su}, \bibinfo{person}{Yuntao Wang},
  \bibinfo{person}{Qichao Xu}, {and} \bibinfo{person}{Ning Zhang}.}
  \bibinfo{year}{2022}\natexlab{}.
\newblock \showarticletitle{{LVBS: Lightweight Vehicular Blockchain for Secure
  Data Sharing in Disaster Rescue}}.
\newblock \bibinfo{journal}{\emph{IEEE Transactions on Dependable and Secure
  Computing}} \bibinfo{volume}{19}, \bibinfo{number}{1} (\bibinfo{year}{2022}),
  \bibinfo{pages}{19--32}.
\newblock
\urldef\tempurl%
\url{https://doi.org/10.1109/TDSC.2020.2980255}
\showDOI{\tempurl}


\bibitem[Sun et~al\mbox{.}(2021)]%
        {9565851}
\bibfield{author}{\bibinfo{person}{Peng Sun}, \bibinfo{person}{Haoxuan Che},
  \bibinfo{person}{Zhibo Wang}, \bibinfo{person}{Yuwei Wang},
  \bibinfo{person}{Tao Wang}, \bibinfo{person}{Liantao Wu}, {and}
  \bibinfo{person}{Huajie Shao}.} \bibinfo{year}{2021}\natexlab{}.
\newblock \showarticletitle{{Pain-FL: Personalized Privacy-Preserving Incentive
  for Federated Learning}}.
\newblock \bibinfo{journal}{\emph{IEEE Journal on Selected Areas in
  Communications}} \bibinfo{volume}{39}, \bibinfo{number}{12}
  (\bibinfo{year}{2021}), \bibinfo{pages}{3805--3820}.
\newblock
\urldef\tempurl%
\url{https://doi.org/10.1109/JSAC.2021.3118354}
\showDOI{\tempurl}


\bibitem[Tairi et~al\mbox{.}(2021)]%
        {Tairi2021A2L}
\bibfield{author}{\bibinfo{person}{Erkan Tairi}, \bibinfo{person}{Pedro
  Moreno-Sanchez}, {and} \bibinfo{person}{Matteo Maffei}.}
  \bibinfo{year}{2021}\natexlab{}.
\newblock \showarticletitle{{A2L: Anonymous Atomic Locks for Scalability in
  Payment Channel Hubs}}. In \bibinfo{booktitle}{\emph{IEEE Symposium on
  Security and Privacy (SP)}}. \bibinfo{pages}{1834--1851}.
\newblock
\urldef\tempurl%
\url{https://doi.org/10.1109/SP40001.2021.00111}
\showDOI{\tempurl}


\bibitem[Tang et~al\mbox{.}(2022)]%
        {9918062}
\bibfield{author}{\bibinfo{person}{Fengxiao Tang}, \bibinfo{person}{Cong Wen},
  \bibinfo{person}{Linfeng Luo}, \bibinfo{person}{Ming Zhao}, {and}
  \bibinfo{person}{Nei Kato}.} \bibinfo{year}{2022}\natexlab{}.
\newblock \showarticletitle{{Blockchain-Based Trusted Traffic Offloading in
  Space-Air-Ground Integrated Networks (SAGIN): A Federated Reinforcement
  Learning Approach}}.
\newblock \bibinfo{journal}{\emph{IEEE Journal on Selected Areas in
  Communications}} \bibinfo{volume}{40}, \bibinfo{number}{12}
  (\bibinfo{year}{2022}), \bibinfo{pages}{3501--3516}.
\newblock
\urldef\tempurl%
\url{https://doi.org/10.1109/JSAC.2022.3213317}
\showDOI{\tempurl}


\bibitem[Wan et~al\mbox{.}(2023)]%
        {Wan2022zk}
\bibfield{author}{\bibinfo{person}{Zhiguo Wan}, \bibinfo{person}{Yan Zhou},
  {and} \bibinfo{person}{Kui Ren}.} \bibinfo{year}{2023}\natexlab{}.
\newblock \showarticletitle{{zk-AuthFeed: Protecting Data Feed to Smart
  Contracts With Authenticated Zero Knowledge Proof}}.
\newblock \bibinfo{journal}{\emph{IEEE Transactions on Dependable and Secure
  Computing}} \bibinfo{volume}{20}, \bibinfo{number}{2} (\bibinfo{year}{2023}),
  \bibinfo{pages}{1335--1347}.
\newblock
\urldef\tempurl%
\url{https://doi.org/10.1109/TDSC.2022.3153084}
\showDOI{\tempurl}


\bibitem[Wang et~al\mbox{.}(2019b)]%
        {PengW2019Flash}
\bibfield{author}{\bibinfo{person}{Peng Wang}, \bibinfo{person}{Hong Xu},
  \bibinfo{person}{Xin Jin}, {and} \bibinfo{person}{Tao Wang}.}
  \bibinfo{year}{2019}\natexlab{b}.
\newblock \showarticletitle{{Flash: Efficient Dynamic Routing for Offchain
  Networks}}. In \bibinfo{booktitle}{\emph{International Conference on Emerging
  Networking Experiments and Technologies}}. \bibinfo{pages}{370--381}.
\newblock


\bibitem[Wang et~al\mbox{.}(2019a)]%
        {Wang2019Adaptive}
\bibfield{author}{\bibinfo{person}{Shiqiang Wang}, \bibinfo{person}{Tiffany
  Tuor}, \bibinfo{person}{Theodoros Salonidis}, \bibinfo{person}{Kin~K. Leung},
  \bibinfo{person}{Christian Makaya}, \bibinfo{person}{Ting He}, {and}
  \bibinfo{person}{Kevin Chan}.} \bibinfo{year}{2019}\natexlab{a}.
\newblock \showarticletitle{{Adaptive Federated Learning in Resource
  Constrained Edge Computing Systems}}.
\newblock \bibinfo{journal}{\emph{IEEE Journal on Selected Areas in
  Communications}} \bibinfo{volume}{37}, \bibinfo{number}{6}
  (\bibinfo{year}{2019}), \bibinfo{pages}{1205--1221}.
\newblock
\urldef\tempurl%
\url{https://doi.org/10.1109/JSAC.2019.2904348}
\showDOI{\tempurl}


\bibitem[Wang et~al\mbox{.}(2022a)]%
        {Wang2022DPCN}
\bibfield{author}{\bibinfo{person}{Wenhui Wang}, \bibinfo{person}{Ke Mu}, {and}
  \bibinfo{person}{Xuetao Wei}.} \bibinfo{year}{2022}\natexlab{a}.
\newblock \showarticletitle{{DPCN: Towards Deadline-Aware Payment Channel
  Networks}}.
\newblock \bibinfo{journal}{\emph{arXiv preprint arXiv:2209.10299}}
  (\bibinfo{year}{2022}).
\newblock


\bibitem[Wang et~al\mbox{.}(2023a)]%
        {Wang2023AI}
\bibfield{author}{\bibinfo{person}{Xiaoding Wang}, \bibinfo{person}{Wenxin
  Liu}, \bibinfo{person}{Hui Lin}, \bibinfo{person}{Jia Hu},
  \bibinfo{person}{Kuljeet Kaur}, {and} \bibinfo{person}{M.~Shamim Hossain}.}
  \bibinfo{year}{2023}\natexlab{a}.
\newblock \showarticletitle{{AI-Empowered Trajectory Anomaly Detection for
  Intelligent Transportation Systems: A Hierarchical Federated Learning
  Approach}}.
\newblock \bibinfo{journal}{\emph{IEEE Transactions on Intelligent
  Transportation Systems}} \bibinfo{volume}{24}, \bibinfo{number}{4}
  (\bibinfo{year}{2023}), \bibinfo{pages}{4631--4640}.
\newblock
\urldef\tempurl%
\url{https://doi.org/10.1109/TITS.2022.3209903}
\showDOI{\tempurl}


\bibitem[Wang et~al\mbox{.}(2022b)]%
        {Wang2022Mean}
\bibfield{author}{\bibinfo{person}{Xiaojie Wang}, \bibinfo{person}{Zhaolong
  Ning}, \bibinfo{person}{Lei Guo}, \bibinfo{person}{Song Guo},
  \bibinfo{person}{Xinbo Gao}, {and} \bibinfo{person}{Guoyin Wang}.}
  \bibinfo{year}{2022}\natexlab{b}.
\newblock \showarticletitle{{Mean-Field Learning for Edge Computing in Mobile
  Blockchain Networks}}.
\newblock \bibinfo{journal}{\emph{IEEE Transactions on Mobile Computing,
  DOI:{10.1109/TMC.2022.3186699}}} (\bibinfo{year}{2022}),
  \bibinfo{pages}{1--17}.
\newblock


\bibitem[Wang et~al\mbox{.}(2023b)]%
        {Wang2023Blockchain}
\bibfield{author}{\bibinfo{person}{Xiaojie Wang}, \bibinfo{person}{Hailin Zhu},
  \bibinfo{person}{Zhaolong Ning}, \bibinfo{person}{Lei Guo}, {and}
  \bibinfo{person}{Yan Zhang}.} \bibinfo{year}{2023}\natexlab{b}.
\newblock \showarticletitle{{Blockchain Intelligence for Internet of Vehicles:
  Challenges and Solutions}}.
\newblock \bibinfo{journal}{\emph{IEEE Communications Surveys \& Tutorials,
  DOI:{10.1109/COMST.2023.3305312}}} (\bibinfo{year}{2023}),
  \bibinfo{pages}{1--1}.
\newblock


\bibitem[Wang et~al\mbox{.}(2022c)]%
        {9928220}
\bibfield{author}{\bibinfo{person}{Yuntao Wang}, \bibinfo{person}{Haixia Peng},
  \bibinfo{person}{Zhou Su}, \bibinfo{person}{Tom~H. Luan},
  \bibinfo{person}{Abderrahim Benslimane}, {and} \bibinfo{person}{Yuan Wu}.}
  \bibinfo{year}{2022}\natexlab{c}.
\newblock \showarticletitle{{A Platform-Free Proof of Federated Learning
  Consensus Mechanism for Sustainable Blockchains}}.
\newblock \bibinfo{journal}{\emph{IEEE Journal on Selected Areas in
  Communications}} \bibinfo{volume}{40}, \bibinfo{number}{12}
  (\bibinfo{year}{2022}), \bibinfo{pages}{3305--3324}.
\newblock
\urldef\tempurl%
\url{https://doi.org/10.1109/JSAC.2022.3213347}
\showDOI{\tempurl}


\bibitem[Wang et~al\mbox{.}(2021)]%
        {wang2021lifesaving}
\bibfield{author}{\bibinfo{person}{Yuntao Wang}, \bibinfo{person}{Zhou Su},
  \bibinfo{person}{Qichao Xu}, \bibinfo{person}{Ruidong Li}, {and}
  \bibinfo{person}{Tom~H. Luan}.} \bibinfo{year}{2021}\natexlab{}.
\newblock \showarticletitle{{Lifesaving with RescueChain: Energy-Efficient and
  Partition-Tolerant Blockchain Based Secure Information Sharing for UAV-Aided
  Disaster Rescue}}. In \bibinfo{booktitle}{\emph{IEEE Conference on Computer
  Communications (INFOCOM)}}. \bibinfo{pages}{1--10}.
\newblock
\urldef\tempurl%
\url{https://doi.org/10.1109/INFOCOM42981.2021.9488719}
\showDOI{\tempurl}


\bibitem[Weng et~al\mbox{.}(2022)]%
        {Weng2022Golden}
\bibfield{author}{\bibinfo{person}{Jiasi Weng}, \bibinfo{person}{Jian Weng},
  \bibinfo{person}{Chengjun Cai}, \bibinfo{person}{Hongwei Huang}, {and}
  \bibinfo{person}{Cong Wang}.} \bibinfo{year}{2022}\natexlab{}.
\newblock \showarticletitle{{Golden Grain: Building a Secure and Decentralized
  Model Marketplace for MLaaS}}.
\newblock \bibinfo{journal}{\emph{IEEE Transactions on Dependable and Secure
  Computing}} \bibinfo{volume}{19}, \bibinfo{number}{5} (\bibinfo{year}{2022}),
  \bibinfo{pages}{3149--3167}.
\newblock
\urldef\tempurl%
\url{https://doi.org/10.1109/TDSC.2021.3085988}
\showDOI{\tempurl}


\bibitem[Wu et~al\mbox{.}(2022)]%
        {Wu2022Blockchain}
\bibfield{author}{\bibinfo{person}{Guangjun Wu}, \bibinfo{person}{Shupeng
  Wang}, \bibinfo{person}{Zhaolong Ning}, {and} \bibinfo{person}{Jun Li}.}
  \bibinfo{year}{2022}\natexlab{}.
\newblock \showarticletitle{{Blockchain-Enabled Privacy-Preserving Access
  Control for Data Publishing and Sharing in the Internet of Medical Things}}.
\newblock \bibinfo{journal}{\emph{IEEE Internet of Things Journal}}
  \bibinfo{volume}{9}, \bibinfo{number}{11} (\bibinfo{year}{2022}),
  \bibinfo{pages}{8091--8104}.
\newblock
\urldef\tempurl%
\url{https://doi.org/10.1109/JIOT.2021.3138104}
\showDOI{\tempurl}


\bibitem[Xiao et~al\mbox{.}(2020)]%
        {Xiao2020A}
\bibfield{author}{\bibinfo{person}{Yang Xiao}, \bibinfo{person}{Ning Zhang},
  \bibinfo{person}{Wenjing Lou}, {and} \bibinfo{person}{Y.~Thomas Hou}.}
  \bibinfo{year}{2020}\natexlab{}.
\newblock \showarticletitle{{A Survey of Distributed Consensus Protocols for
  Blockchain Networks}}.
\newblock \bibinfo{journal}{\emph{IEEE Communications Surveys \& Tutorials}}
  \bibinfo{volume}{22}, \bibinfo{number}{2} (\bibinfo{year}{2020}),
  \bibinfo{pages}{1432--1465}.
\newblock
\urldef\tempurl%
\url{https://doi.org/10.1109/COMST.2020.2969706}
\showDOI{\tempurl}


\bibitem[Xie et~al\mbox{.}(2022)]%
        {xie2022sofitmix}
\bibfield{author}{\bibinfo{person}{Haomeng Xie}, \bibinfo{person}{Shufan Fei},
  \bibinfo{person}{Zheng Yan}, {and} \bibinfo{person}{Yang Xiao}.}
  \bibinfo{year}{2022}\natexlab{}.
\newblock \showarticletitle{{SofitMix: A Secure Offchain-Supported
  Bitcoin-Compatible Mixing Protocol}}.
\newblock \bibinfo{journal}{\emph{IEEE Transactions on Dependable and Secure
  Computing, DOI:{10.1109/TDSC.2022.3213824}}} (\bibinfo{year}{2022}),
  \bibinfo{pages}{1--15}.
\newblock


\bibitem[Xie et~al\mbox{.}(2019)]%
        {Xie2019A}
\bibfield{author}{\bibinfo{person}{Junfeng Xie}, \bibinfo{person}{Helen Tang},
  \bibinfo{person}{Tao Huang}, \bibinfo{person}{F.~Richard Yu},
  \bibinfo{person}{Renchao Xie}, \bibinfo{person}{Jiang Liu}, {and}
  \bibinfo{person}{Yunjie Liu}.} \bibinfo{year}{2019}\natexlab{}.
\newblock \showarticletitle{{A Survey of Blockchain Technology Applied to Smart
  Cities: Research Issues and Challenges}}.
\newblock \bibinfo{journal}{\emph{IEEE Communications Surveys \& Tutorials}}
  \bibinfo{volume}{21}, \bibinfo{number}{3} (\bibinfo{year}{2019}),
  \bibinfo{pages}{2794--2830}.
\newblock
\urldef\tempurl%
\url{https://doi.org/10.1109/COMST.2019.2899617}
\showDOI{\tempurl}


\bibitem[Xue et~al\mbox{.}(2022)]%
        {Xue2022Blockchain}
\bibfield{author}{\bibinfo{person}{Liang Xue}, \bibinfo{person}{Dongxiao Liu},
  \bibinfo{person}{Cheng Huang}, \bibinfo{person}{Xuemin Shen},
  \bibinfo{person}{Weihua Zhuang}, \bibinfo{person}{Rob Sun}, {and}
  \bibinfo{person}{Bidi Ying}.} \bibinfo{year}{2022}\natexlab{}.
\newblock \showarticletitle{{Blockchain-Based Data Sharing With Key Update for
  Future Networks}}.
\newblock \bibinfo{journal}{\emph{IEEE Journal on Selected Areas in
  Communications}} \bibinfo{volume}{40}, \bibinfo{number}{12}
  (\bibinfo{year}{2022}), \bibinfo{pages}{3437--3451}.
\newblock


\bibitem[Ye et~al\mbox{.}(2022)]%
        {Ye2022FLasaS}
\bibfield{author}{\bibinfo{person}{Wenxuan Ye}, \bibinfo{person}{Xueli Ant},
  \bibinfo{person}{Xueqiang Yan}, \bibinfo{person}{Mohammad Hamad}, {and}
  \bibinfo{person}{Sebastian Steinhorst}.} \bibinfo{year}{2022}\natexlab{}.
\newblock \showarticletitle{{FLaaS6G: Federated Learning as a Service in 6G
  Using Distributed Data Management Architecture}}. In
  \bibinfo{booktitle}{\emph{IEEE Global Communications Conference (GLOBECOM)}}.
  \bibinfo{pages}{1247--1252}.
\newblock
\urldef\tempurl%
\url{https://doi.org/10.1109/GLOBECOM48099.2022.10001307}
\showDOI{\tempurl}


\bibitem[Yu et~al\mbox{.}(2018)]%
        {yu2018coinexpress}
\bibfield{author}{\bibinfo{person}{Ruozhou Yu}, \bibinfo{person}{Guoliang Xue},
  \bibinfo{person}{Vishnu~Teja Kilari}, \bibinfo{person}{Dejun Yang}, {and}
  \bibinfo{person}{Jian Tang}.} \bibinfo{year}{2018}\natexlab{}.
\newblock \showarticletitle{{CoinExpress: A Fast Payment Routing Mechanism in
  Blockchain-Based Payment Channel Networks}}. In
  \bibinfo{booktitle}{\emph{International Conference on Computer Communication
  and Networks (ICCCN)}}. \bibinfo{pages}{1--9}.
\newblock
\urldef\tempurl%
\url{https://doi.org/10.1109/ICCCN.2018.8487351}
\showDOI{\tempurl}


\bibitem[Yuan et~al\mbox{.}(2023)]%
        {Yuan2023TRUCON}
\bibfield{author}{\bibinfo{person}{Mingyang Yuan}, \bibinfo{person}{Yang Xu},
  \bibinfo{person}{Cheng Zhang}, \bibinfo{person}{Yunlin Tan},
  \bibinfo{person}{Yichuan Wang}, \bibinfo{person}{Ju Ren}, {and}
  \bibinfo{person}{Yaoxue Zhang}.} \bibinfo{year}{2023}\natexlab{}.
\newblock \showarticletitle{{TRUCON: Blockchain-Based Trusted Data Sharing With
  Congestion Control in Internet of Vehicles}}.
\newblock \bibinfo{journal}{\emph{IEEE Transactions on Intelligent
  Transportation Systems}} \bibinfo{volume}{24}, \bibinfo{number}{3}
  (\bibinfo{year}{2023}), \bibinfo{pages}{3489--3500}.
\newblock
\urldef\tempurl%
\url{https://doi.org/10.1109/TITS.2022.3226500}
\showDOI{\tempurl}


\bibitem[Yue et~al\mbox{.}(2021)]%
        {Yue2021A}
\bibfield{author}{\bibinfo{person}{Kaifeng Yue}, \bibinfo{person}{Yuanyuan
  Zhang}, \bibinfo{person}{Yanru Chen}, \bibinfo{person}{Yang Li},
  \bibinfo{person}{Lian Zhao}, \bibinfo{person}{Chunming Rong}, {and}
  \bibinfo{person}{Liangyin Chen}.} \bibinfo{year}{2021}\natexlab{}.
\newblock \showarticletitle{{A Survey of Decentralizing Applications via
  Blockchain: The 5G and Beyond Perspective}}.
\newblock \bibinfo{journal}{\emph{IEEE Communications Surveys \& Tutorials}}
  \bibinfo{volume}{23}, \bibinfo{number}{4} (\bibinfo{year}{2021}),
  \bibinfo{pages}{2191--2217}.
\newblock
\urldef\tempurl%
\url{https://doi.org/10.1109/COMST.2021.3115797}
\showDOI{\tempurl}


\bibitem[Zhang et~al\mbox{.}(2020b)]%
        {Zhang2018Anonymous}
\bibfield{author}{\bibinfo{person}{Di Zhang}, \bibinfo{person}{Junqing Le},
  \bibinfo{person}{Nankun Mu}, {and} \bibinfo{person}{Xiaofeng Liao}.}
  \bibinfo{year}{2020}\natexlab{b}.
\newblock \showarticletitle{{An Anonymous Off-Blockchain Micropayments Scheme
  for Cryptocurrencies in the Real World}}.
\newblock \bibinfo{journal}{\emph{IEEE Transactions on SMC: Systems}}
  \bibinfo{volume}{50}, \bibinfo{number}{1} (\bibinfo{year}{2020}),
  \bibinfo{pages}{32--42}.
\newblock
\urldef\tempurl%
\url{https://doi.org/10.1109/TSMC.2018.2884289}
\showDOI{\tempurl}


\bibitem[Zhang et~al\mbox{.}(2023b)]%
        {Zhang2021Boros}
\bibfield{author}{\bibinfo{person}{Jingjing Zhang}, \bibinfo{person}{Yongjie
  Ye}, \bibinfo{person}{Weigang Wu}, {and} \bibinfo{person}{Xiapu Luo}.}
  \bibinfo{year}{2023}\natexlab{b}.
\newblock \showarticletitle{{Boros: Secure and Efficient Off-Blockchain
  Transactions via Payment Channel Hub}}.
\newblock \bibinfo{journal}{\emph{IEEE Transactions on Dependable and Secure
  Computing}} \bibinfo{volume}{20}, \bibinfo{number}{1} (\bibinfo{year}{2023}),
  \bibinfo{pages}{407--421}.
\newblock
\urldef\tempurl%
\url{https://doi.org/10.1109/TDSC.2021.3135076}
\showDOI{\tempurl}


\bibitem[Zhang et~al\mbox{.}(2020a)]%
        {zhang2020off}
\bibfield{author}{\bibinfo{person}{Lei Zhang}, \bibinfo{person}{Sanjay Bakshi},
  {and} \bibinfo{person}{John~K Zao}.} \bibinfo{year}{2020}\natexlab{a}.
\newblock \showarticletitle{{Off-Chain Trusted Computing}}.
\newblock \bibinfo{journal}{\emph{IEEE Internet of Things Magazine}}
  \bibinfo{volume}{3}, \bibinfo{number}{2} (\bibinfo{year}{2020}),
  \bibinfo{pages}{8--9}.
\newblock


\bibitem[Zhang et~al\mbox{.}(2022)]%
        {zhang2022blockchain}
\bibfield{author}{\bibinfo{person}{Yue Zhang}, \bibinfo{person}{Keke Gai},
  \bibinfo{person}{Jiang Xiao}, \bibinfo{person}{Liehuang Zhu}, {and}
  \bibinfo{person}{Kim-Kwang~Raymond Choo}.} \bibinfo{year}{2022}\natexlab{}.
\newblock \showarticletitle{{Blockchain-Empowered Efficient Data Sharing in
  Internet of Things Settings}}.
\newblock \bibinfo{journal}{\emph{IEEE Journal on Selected Areas in
  Communications}} \bibinfo{volume}{40}, \bibinfo{number}{12}
  (\bibinfo{year}{2022}), \bibinfo{pages}{3422--3436}.
\newblock
\urldef\tempurl%
\url{https://doi.org/10.1109/JSAC.2022.3213353}
\showDOI{\tempurl}


\bibitem[Zhang et~al\mbox{.}(2023a)]%
        {Zhang2023Anonymous}
\bibfield{author}{\bibinfo{person}{Yi Zhang}, \bibinfo{person}{Xiaofeng Jia},
  \bibinfo{person}{Bianjing Pan}, \bibinfo{person}{Jun Shao},
  \bibinfo{person}{Liming Fang}, \bibinfo{person}{Rongxing Lu}, {and}
  \bibinfo{person}{Guiyi Wei}.} \bibinfo{year}{2023}\natexlab{a}.
\newblock \showarticletitle{{Anonymous Multi-Hop Payment for Payment Channel
  Networks}}.
\newblock \bibinfo{journal}{\emph{IEEE Transactions on Dependable and Secure
  Computing, DOI:{10.1109/TDSC.2023.3262681}}} (\bibinfo{year}{2023}),
  \bibinfo{pages}{1--11}.
\newblock


\bibitem[Zhang et~al\mbox{.}(2016)]%
        {Zhang2016Social}
\bibfield{author}{\bibinfo{person}{Yue Zhang}, \bibinfo{person}{Fang Tian},
  \bibinfo{person}{Bin Song}, {and} \bibinfo{person}{Xiaojiang Du}.}
  \bibinfo{year}{2016}\natexlab{}.
\newblock \showarticletitle{{Social Vehicle Swarms: A Novel Perspective on
  Socially Aware Vehicular Communication Architecture}}.
\newblock \bibinfo{journal}{\emph{IEEE Wireless Communications}}
  \bibinfo{volume}{23}, \bibinfo{number}{4} (\bibinfo{year}{2016}),
  \bibinfo{pages}{82--89}.
\newblock
\urldef\tempurl%
\url{https://doi.org/10.1109/MWC.2016.7553030}
\showDOI{\tempurl}


\bibitem[Zhang and Yang(2019)]%
        {Zhang2021Robustpay}
\bibfield{author}{\bibinfo{person}{Yuhui Zhang} {and} \bibinfo{person}{Dejun
  Yang}.} \bibinfo{year}{2019}\natexlab{}.
\newblock \showarticletitle{{RobustPay: Robust Payment Routing Protocol in
  Blockchain-Based Payment Channel Networks}}. In
  \bibinfo{booktitle}{\emph{IEEE International Conference on Network Protocols
  (ICNP)}}. \bibinfo{pages}{1--4}.
\newblock


\bibitem[Zhang and Yang(2021)]%
        {Zhang2021RobustPay+}
\bibfield{author}{\bibinfo{person}{Yuhui Zhang} {and} \bibinfo{person}{Dejun
  Yang}.} \bibinfo{year}{2021}\natexlab{}.
\newblock \showarticletitle{{RobustPay+: Robust Payment Routing With
  Approximation Guarantee in Blockchain-Based Payment Channel Networks}}.
\newblock \bibinfo{journal}{\emph{IEEE/ACM Transactions on Networking}}
  \bibinfo{volume}{29}, \bibinfo{number}{4} (\bibinfo{year}{2021}),
  \bibinfo{pages}{1676--1686}.
\newblock
\urldef\tempurl%
\url{https://doi.org/10.1109/TNET.2021.3069725}
\showDOI{\tempurl}


\bibitem[Zhao et~al\mbox{.}(2022)]%
        {Zhao2022ATensor}
\bibfield{author}{\bibinfo{person}{Ruonan Zhao}, \bibinfo{person}{Laurence~T.
  Yang}, \bibinfo{person}{Debin Liu}, \bibinfo{person}{Xianjun Deng}, {and}
  \bibinfo{person}{Yijun Mo}.} \bibinfo{year}{2022}\natexlab{}.
\newblock \showarticletitle{{A Tensor-Based Truthful Incentive Mechanism for
  Blockchain-Enabled Space-Air-Ground Integrated Vehicular Crowdsensing}}.
\newblock \bibinfo{journal}{\emph{IEEE Transactions on Intelligent
  Transportation Systems}} \bibinfo{volume}{23}, \bibinfo{number}{3}
  (\bibinfo{year}{2022}), \bibinfo{pages}{2853--2862}.
\newblock
\urldef\tempurl%
\url{https://doi.org/10.1109/TITS.2022.3144301}
\showDOI{\tempurl}


\bibitem[Zheng et~al\mbox{.}(2020)]%
        {Zheng2017Privacy}
\bibfield{author}{\bibinfo{person}{Yifeng Zheng}, \bibinfo{person}{Huayi Duan},
  \bibinfo{person}{Xingliang Yuan}, {and} \bibinfo{person}{Cong Wang}.}
  \bibinfo{year}{2020}\natexlab{}.
\newblock \showarticletitle{{Privacy-Aware and Efficient Mobile Crowdsensing
  with Truth Discovery}}.
\newblock \bibinfo{journal}{\emph{IEEE Transactions on Dependable and Secure
  Computing}} \bibinfo{volume}{17}, \bibinfo{number}{1} (\bibinfo{year}{2020}),
  \bibinfo{pages}{121--133}.
\newblock
\urldef\tempurl%
\url{https://doi.org/10.1109/TDSC.2017.2753245}
\showDOI{\tempurl}


\end{thebibliography}

\end{document}